\documentclass[preprint]{aastex}
\usepackage{natbib}
 \usepackage[rgb]{xcolor}
\shorttitle{Magnetometry of Superpenumbral Fibrils}
\shortauthors{Schad et al.}
\begin{document}
\newcommand{\refasen}{AR08}

\title{\ion{He}{1} Vector Magnetometry of Field Aligned Superpenumbral Fibrils}

\author{T. A. Schad\altaffilmark{1}}
\affil{Department of Planetary Sciences, University of Arizona, Tucson, AZ 85721}
\author{M.J. Penn}
\affil{National Solar Observatory, Tucson, AZ 85719}
\author{H. Lin}
\affil{Institute for Astronomy, University of Hawai'i, Pukalani, HI 96768}

\altaffiltext{1}{Graduate student researcher at the National Solar Observatory (NSO), operated by the Association of Universities for Research in Astronomy, Inc. (AURA), under cooperative agreement with the National Science Foundation.}


\begin{abstract}
Atomic-level polarization and Zeeman effect diagnostics in the neutral helium triplet at 10830 \mbox{\AA} in principle allow full vector magnetometry of fine-scaled chromospheric fibrils.  We present high-resolution spectropolarimetric observations of superpenumbral fibrils in the \ion{He}{1} triplet with sufficient polarimetric sensitivity to infer their full magnetic field geometry.  \ion{He}{1} observations from the Facility Infrared Spectropolarimeter (FIRS) are paired with high-resolution observations of the H$\alpha$ 6563 \mbox{\AA} and \ion{Ca}{2} 8542 \mbox{\AA} spectral lines from the Interferometric Bidimensional Spectrometer (IBIS) from the Dunn Solar Telescope in New Mexico.  Linear and circular polarization signatures in the \ion{He}{1} triplet are measured and described, as well as analyzed with the advanced inversion capability of the ``Hanle and Zeeman Light" (HAZEL) modeling code.  Our analysis provides direct evidence for the often assumed field alignment of fibril structures.  The projected angle of the fibrils and the inferred magnetic field geometry align within an error of $\pm 10^{\circ}$.  We describe changes in the inclination angle of these features that reflect their connectivity with the photospheric magnetic field.  Evidence for an accelerated flow ($\sim 40$ m sec$^{-2}$) along an individual fibril anchored at its endpoints in the strong sunspot and weaker plage in part supports the magnetic siphon-flow mechanism's role in the inverse Evershed effect.  However, the connectivity of the outer endpoint of many of the fibrils cannot be established.  
\end{abstract}



\section{Introduction}\label{sec:intro} 

Extending laterally away from magnetic field concentrations in the solar photosphere, threadlike fibrils observed in the chromosphere \citep[e.g.][]{hale1908,veeder1970,pietarila2009,reardon2011} host a variety of dynamic behavior and have long been considered tracers of the difficult to measure chromospheric magnetic field \citep{smith1968,foukal1971_2,foukal1971_1}.  Outside of sunspots, fibrils extend across internetwork cells and often, but not always, visibly suggest that their endpoints are rooted in areas of opposite polarity photospheric network flux.  For this reason, fibrils are thought to be field aligned closed magnetic loop segments.  In the case of fibrils with endpoints not associated with opposite polarity network flux, \cite{foukal1971_1} suggest that H$\alpha$ fibrils may extend over the neighboring flux and connect with regions of opposite polarity flux at a further distance, and thus the fibrils being directed to higher heights disappear in H$\alpha$.  Contrarily, \cite{reardon2011}, using high-resolution \ion{Ca}{2} 8542 \mbox{\AA} narrowband images, argues that such internetwork fibrils are connected to the weak internetwork field directly below, meaning the fibrils do not contain much of the total flux of the concentrations from which they extend.  The remaining flux would be directed upwards into the corona, and have direct influence on dynamic heating mechanisms.

Similarly, fibrils and/or threads surrounding sunspots form what is known as the superpenumbra \citep{loughhead1968}.  Again these fibrils are presumably magnetic loops rooted at one end in the sunspot and at the other in some opposite polarity plage.  This concept is invoked by the siphon flow model to explain the apparent inward flow (i.e. the Inverse Evershed effect) directed along the superpenumbral fibrils \citep{evershed1909,meyer1968,maltby1975}.  Yet, characteristic shocks thought to be formed by siphon flows have not been definitively observed; though, some evidence does exist \citep{uitenbroek2006,bethge2012}.  Other short-lived phenomena are also witnessed in superpenumbral fibrils that are difficult to explain in terms of simple siphon flow \citep{vissers2012}.

Until recently our knowledge of the fine-scaled fibril magnetic field had been limited to the morphological constraints placed on fine structure via comparison with photospheric field measurements. \cite{wiegelmann2008} noticed non-linear force free extrapolations of the photospheric magnetic field better account for the free-magnetic energy of the corona when the fibril direction is used as a constraint on the horizontal field direction; yet, direct measurement remains lacking.  Moreover, the photospheric field is a poor boundary condition since it is not strictly force-free.  \cite{lagg2009} studied two curvilinear internetwork structures observed with the \ion{He}{1} triplet at 10830 \mbox{\AA} and found evidence that the structures hosted a horizontal magnetic field, but the primary fibril axis was offset from the inferred magnetic field direction in one of the structures.  To this add that \cite{asensio_ramos2008} (hereafter \refasen) interpreted unresolved disk center \ion{He}{1} internetwork spectra and found less inclined fields ($\theta_B$ $\approx$ $21^{\circ}$ or $\theta_B$ $\approx$ $47^{\circ}$ w.r.t. solar vertical).  More recently, \cite{delaCruz2011} studied the transverse Zeeman effect on the \ion{Ca}{2} 8542 \mbox{\AA} spectral line within superpenumbral fibrils near/above the external boundary of a photospheric penumbra and evidenced general consistency between the fibril axes and the inferred transverse magnetic field direction.  They, however, pointed out particular cases where the inferred field did not match the visible morphology.  Furthermore, the \ion{Ca}{2} linear polarization decreased rapidly outwards from the sunspot and was considered inconsistent with the presumed superpenumbral canopy. 

We address the magnetic field vector within resolved chromospheric fibrils using high-resolution observations of the \ion{He}{1} infrared triplet.  Although the utility of these three spectral lines has been long emphasized as indicators of solar and stellar activity \citep{zirin1982,kozlova2003} and probes of the chromospheric magnetic field \citep{harvey1971,ruedi1995}, only recently has the theoretical framework of its polarized spectral line formation (\cite{trujillo_bueno_2007}; \refasen) been met in maturity by instruments capable of measuring its weak polarization signals at spatial scales of interest (see, e.g., \cite{collados2007} and \cite{jaeggli2010}).  Here we discuss the polarization signatures of the \ion{He}{1} triplet observed within an active region superpenumbra observed with multi-channel instrumentation at the Dunn Solar Telescope (\S\ref{sec:obs_reduce}, \S\ref{sec:multi_wave}).  A heuristic description of the mechanisms inducing \ion{He}{1} polarization is given in \S \ref{sec:spectra_description} to explain the macroscopic structure seen in polarized maps of the active region. Section~\ref{sec:methods} outlines our methods to model the visible fibril morphology as well as the inversion method used to infer the magnetic field vector from the polarized spectra.  A summary and discussion of the results follow. 


\section{Observations and Data Reduction}\label{sec:obs_reduce}

On 2012 January 29, NOAA active region (AR) 11408 consisted of a simple alpha-type sunspot in a bipolar configuration with trailing plage.  We targeted this region at the National Solar Observatory's (NSO) Dunn Solar Telescope (DST) located on Sacramento Peak in New Mexico, USA.  The observations included multi-channel, imaging and slit-type, spectroscopy and spectropolarimetry using the Interferometric BiDimensional Spectrometer \citep[IBIS:][]{cavallini2006,reardon2008} and the Facility Infrared Spectropolarimeter \citep[FIRS:][]{jaeggli2010}.  An additional camera acquired broadband images within the 4300 \mbox{\AA} molecular spectral G-band.  All instruments were operated simultaneously and benefited from the facility's High Order Adaptive Optics system \citep[HOAO:][]{rimmele2004} during a period of good to excellent seeing. 

\begin{figure*}
\epsscale{1.}
\plotone{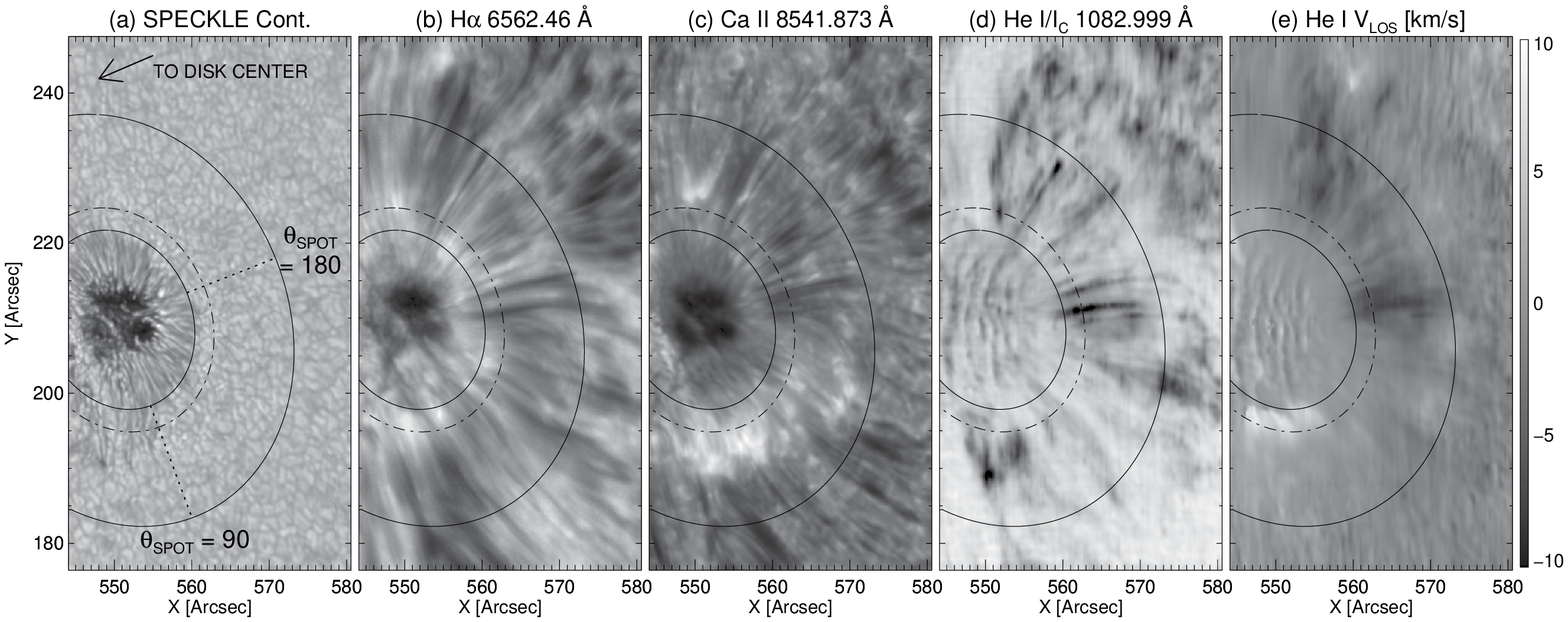}
\caption{NOAA active region 11408 observed by FIRS and IBIS on 2012 January 29. (a) SPECKLE reconstructed broadband images from IBIS at 20:05 UT (b) IBIS H$\alpha$ intensity image at 6562.46 \mbox{\AA} acquired at 20:05:16 UT (c) IBIS \ion{Ca}{2} intensity image at 8541.873 \mbox{\AA} acquired at 20:04:53 UT (d) FIRS map of the \ion{He}{1} relative intensity (i.e. I($\lambda$) normalized to local continuum) at $\lambda = 10829.995$ \mbox{\AA}.  The slit, oriented parallel to the solar meridian, was scanned from solar east to west at a rate of $0.64''$ min$^{-1}$ (e) \ion{He}{1} Doppler velocity corrected for orbital motions and solar rotation.  Solar north is up and the x and y axes give helioprojective coordinates.  \label{fig:fov_maps}}
\end{figure*}

Our focus here is on a single map of NOAA AR 11408 acquired in the \ion{He}{1} triplet with the FIRS dual-beam slit-spectropolarimeter when the AR was located at N8W35 ($\mu = \cos{\Theta} = 0.8$).  A single slit was oriented parallel to the solar north-south axis and scanned across the region from east to west with a projected step size and slit width of $0.3''$\,--\,the DST angular resolution is limited to $0.36''$ at 10830 \mbox{\AA} by the Rayleigh criterion. Seventy-seven (77) arc-seconds were imaged along the slit.  A 200 step map commenced at 19:16 UT and required 94 minutes to complete.  The common IBIS and FIRS FOV is illustrated in Figure~\ref{fig:fov_maps}.  This deep FIRS scan measured the full Stokes state of the incoming light with a 4-state efficiency-balanced modulation scheme with 125 msec exposures.  At each step position 15 consecutive modulation sequences were coadded for a total integration time of 7.5 seconds per slit position. The final spatial sampling after reduction (described below) is $0.3'' \times 0.3''$.   Spectral sampling is $38.6803$ m$\mbox{\AA}$ pix$^{-1}$ over a 35.74 $\mbox{\AA}$ spectral range.  The mean noise level in the Stokes Q, U, and V spectra is  $4.0\times10^{-4}$, $3.7\times10^{-4}$, and $3.0\times10^{-4}$, respectively, in units of continuum intensity.   

In congress with FIRS, high-resolution full Stokes polarimetry was performed with IBIS, a dual Fabry-Perot interferometer, over a  $45'' \times 95''$ field-of-view (FOV).  Full Stokes measurements were acquired in the photospheric Fe I 6173 \mbox{\AA} spectral line at $N_{\lambda} =16$ distributed wavelength points across the line and continuum, as well as in the chromospheric \ion{Ca}{2} 8542 \mbox{\AA} line ($N_{\lambda} = 20$).  In addition, we performed imaging spectroscopy of the H$\alpha$ 6563 \mbox{\AA} line ($N_{\lambda} = 22$).  Normal reduction methods were used (see, e.g., \cite{judge2010}), and included full polarimetric calibration, FOV-dependent spectral wavelength shift correction, and spatial destretching using SPECKLE reconstructed broadband images recorded simultaneously at 6360 \mbox{\AA} \citep{woger2008}.  The entire observation cycle of the three channels lasted 50 seconds and was repeated throughout all other observations.  

To verify the observational geometry needed to specify the scattering angle of the \ion{He}{1} observations, we coaligned all observations with continuum data from the Helioseismic and Magnetic Imager \citep[HMI:][]{scherrer2012} on board NASA's Solar Dynamics Observatory (SDO).  We also employ SDO/HMI magnetograms for potential field extrapolations in Section~\ref{sec:fibril_magnetic_maps}.  The standard preparation routines available for SDO/HMI data, including \textit{aia\_prep} v4.13, were utilized.  All observations from the DST were coregistered and corrected for differential atmospheric refraction.  We estimate the coregistration error is less than $0.5''$.  The largest error in the observed geometry of the FIRS map is the 0.8 degree change in the heliographic observation angle introduced by solar rotation.

\subsection{Reduction of FIRS Spectra}

The reduction of the raw FIRS spectra differs in a number of ways from previous methods \citep{jaeggli2012}.  Detector properties were calibrated via dark and bias subtraction, spectral flat fields, and a non-linearity correction; though, these observations were careful to utilize only the most linear portion of the detector response curve.  Flat fields consisted of averaging quiet sun observations taken at disk center just after the science observations.  The spectral lines are removed in the spectral flats prior to their application on the science data with Voigt profiles fits to each line.  Weak lines are not well removed in this process due to large pixel-to-pixel gain variations.  Deep lines, such as the Si I line at 10827.089$\mbox{\AA}$, are well calibrated.  Within the \ion{He}{1} triplet, the flat-fielding process is quite reliable due to the very weak absorption in the quiet Sun relative to strong signals in the active region which makes its removal in the raw flat-field unnecessary.  

\subsubsection{Polarization Calibration}

Polarization sensitivity and accuracy are critical to our measurements.  As each optical element of the telescope and instrument can modify the state of polarization, we carefully calibrate the polarimetric response of the entire optical system.  Ignoring attenuation and the effects of the detector, the measurement process used by FIRS to derive the polarization state of the incoming solar light can be written as:
\begin{equation}
{\bf{I}_{meas} = \bf{O}\bf{X}\bf{T}\bf{S_{in}},}
\end{equation}
where $\bf{I}_{meas}$ is a vector of four measured intensities.  $\bf{S_{in}}$ is the input Stokes vector, $\bf{T}$ is the collective Mueller matrix of the telescopic optics and is time-dependent due to configuration changes of the telescope throughout the observations.  $\bf{X}$ is the time-independent Mueller matrix of all the instrumental optics following the set of calibration optics that are inserted near the prime telescope focus during calibration.  ${\bf{}X}$ includes the effects of the adaptive optics system, beam splitters, and the FIRS instrument.  $\bf{O}$ is the modulation matrix describing the polarimeter modulation scheme.  

$\bf{T}$ is determined via the sub-aperture polarization calibration scheme described by \cite{socas_navarro2011}.  As in \cite{beck2005} and \cite{socas_navarro2006}, a linear polarizer and wave plate are introduced into the beam to determine $\bf{X}$.  To combat the effects of light-level variations, we fit for the 15 elements of the \textit{normalized} $\bf{X}$ Mueller matrix as in \cite{ichimoto2008}.  We also determine the unknown quantities of the calibration optics, namely the retardance and offset angle of the wave plate, through a least-squares fit to an appropriate calibration optics model.  The linear polarizer is kept in the beam throughout the calibration to decouple the telescope from the downstream optics as in \cite{beck2005}; though, unlike that work, we find the parameters of the wave plate can be found directly from a model fit. 

After demodulation and the correction for $\bf{X}$ and $\bf{T}$, small offsets of the ``unpolarized'' continuum from zero polarization are used to fine-tune the intensity crosstalk calibration \citep{sanchez_almeida1992}.  The two beams of FIRS are then combined to mitigate the effects of seeing-induced crosstalk.  We analyze the residual crosstalk with the correlation method of \cite{schlichenmaier2002} applied to the Si I line at 10827 \mbox{\AA}.  The crosstalk coefficients ${C_{VQ},C_{VU},C_{QV},C_{UV}}$ yield small values of ${0.0198,0.076,-0.013,-0.039}$, which are implemented as a correction.  For the small polarization signals examined in this data (on order of $0.1\%$), the small residual crosstalk coefficients translate to a small polarization error ($0.076 \times 0.1\% =0.0076\%$), smaller than the mean noise level of $\sim 0.03\%$.  Finally, we correct for the time-dependent parallactic angle and rotate the Stokes reference direction such that the Stokes +Q direction is oriented parallel to the solar east/west direction. 

\subsubsection{Polarized Fringe Removal}

Our FIRS observations contain significant polarized fringes that cannot be completely decoupled or removed from the real solar signal by flat-fielding and/or Fourier filtering.  \cite{casini2012} developed a pattern-recognition-based technique using 2D \textit{principal component analysis} (2DPCA) to address this problem, specifically using FIRS data.  We apply these techniques to our data to each beam separately just prior to combination.  A key difference here is that we use the spectral flat fields to determine the projection vectors necessary to ``train'' the 2DPCA algorithm.  In doing so, we reconstruct via the proper selection of eigenfeatures the fringe pattern in the data frames instead of the solar signal itself.  The reconstructed fringes are Fourier filtered using a 2D Fourier analysis, and then detrended, meaning that at each sampled wavelength, the fringe signal is fit as a smooth function of spatial position across the map.  The spatially-detrended, Fourier-filtered, 2DPCA reconstructed fringes are subtracted from the original data frames to recover the true solar signal.  This technique achieves a large increase in the sensitivity of the measurements, suppressing the fringe signal to at or below the photon noise. 

\subsubsection{Wavelength and Velocity Calibration}

Since we are interested in the absolute velocity structure along each fibril, we establish an absolute wavelength scale for these FIRS measurements.   Our observations exhibit only weak telluric absorption on this date, which disallows the use of the method used by \cite{kuckein2012} since the two telluric lines are influenced negatively by the flat-field errors discussed above.  Our wavelength calibration relies instead on the cores of the two deep photospheric Si I lines at 10827.089$\mbox{\AA}$ and 10843.845$\mbox{\AA}$.  The separation of these two wavelengths, which have been corrected for convective blueshift and gravitational redshift by \cite{borrero2003}, is only 1.3 m\mbox{\AA} different than the separation we determine from the FTS spectral atlas \citep{kurucz1984}.  Thus, we argue that this separation can reliably be used to calculate the spectral dispersion.  We find a linear dispersion of 38.6803 m$\mbox{\AA}$/pix, which is consistent with spectral dispersions calculated during other FIRS observing campaigns when the level of telluric absorption was high enough to allow for the \cite{kuckein2012} method.

Since the convective blueshift of the Si I line at 10827.089 $\mbox{\AA}$ is negligible \citep{kuckein2012}, we utilize its observed position for the absolute wavelength calibration.  We assume the line center position of the Si I observed at disk center during the flat fields is shifted by the various orbital motions described by \cite{martinez_pillet_1997} in addition to gravitational redshift.  That is, the average disk-center Doppler velocity for the Si I line after the full velocity calibration is assumed to be zero.  Applying this correction to our science data at a heliographic angle of $\approx 36^{\circ}$, we find the observed Doppler velocities in a patch of quiet sun to be consistent within $\pm 250 $m/s with the determined orbital and gravitational effects, which we take as the error of our Doppler velocities after full correction.


\section{Multi-wavelength Comparison of Chromospheric Fibrils}\label{sec:multi_wave}

A particular strength of these observations is the coordinated diagnostics of H$\alpha$ $6563$ \mbox{\AA}, \ion{Ca}{2} $8542$ \mbox{\AA}, and \ion{He}{1} $10830$ \mbox{\AA} each at sufficient resolution to individuate multiple chromospheric fibrils.  The spectral line formation of each of these lines constrains the thermodynamic structure of the fibrils.  While the formation and dynamics of real solar fibrils is not understood, fibril-like thermodynamic structures are seen within advanced numerical radiative MHD simulations.  Forward radiative transfer (RT) calculations through these simulations are now being used to compare real observed structures with synthetic structures.  Though, the complex non-LTE formation of chromosphere lines remains a challenge.  Fibril structures have not yet been produced in forward RT calculations of the \ion{Ca}{2} 8542 \mbox{\AA} line.  However, recent three-dimensional (3D) non-LTE RT calculations for the H$\alpha$ spectral line through a 3D radiative-MHD simulation including the convective zone and lower corona do produce fibril-like phenomena.  These calculations propose that H$\alpha$ fibrils denote field-aligned ridges of increased mass density at higher average formation heights than the background plasma \citep{leenaarts2012}.

As seen in Figure~\ref{fig:fov_maps}, the fibrils extending outward from the sunspot are remarkably spatially coherent between each spectral line despite the FIRS map being significantly non-monochronic ($t_{scan} = 94$ min) compared to the relatively quick spectral scans ($\delta t < 30$ seconds) of IBIS.  The three center images of Figure~\ref{fig:fov_maps} show the spectral intensity (or relative intensity) in the blue wing of each spectral line.  The fibrils directly associated with the sunspot display the greatest correspondence between the spectral lines, whereas the very fine-scaled fibril features outside of the superpenumbra (near $\left \langle X,Y \right \rangle = \left \langle 570,235 \right \rangle $) are only resolved in the IBIS images.  These fibrils are likely too thin and/or too dynamic to be seen in the lower-resolution FIRS maps.  Individual fibril widths in the superpenumbra (especially at $\left \langle560,230 \right \rangle$ and  $\left \langle 565,210 \right \rangle$) are comparable for each line, suggesting again that these spectral lines probe plasma in the same structures and/or overall topology.  While the temporal stability of the superpenumbral structure as a whole has been previously observed on timescales of hours \citep{loughhead1968}, individual lifetimes of the fibrils are only on the order of tens of minutes \citep{maltby1975}.  It is, however, this stability that allows us to probe these fibrils with \ion{He}{1} 1083 spectropolarimetry since FIRS require long integration times to achieve the necessary sensitivity in the polarized spectra.

The velocity structure in the \ion{He}{1} triplet (Figure~\ref{fig:fov_maps}e) displays the familiar inverse Evershed effect with LOS velocities up to $\sim{ }8-9$ $km$ $sec^{-1}$, which is higher than velocities reported by \cite{penn2002} from lower-resolution observations.  The fibrils themselves exhibit the inverse Evershed effect as well as the inter-fibril material which still contains significant absorption in the \ion{He}{1} triplet.  Chromospheric umbral p-mode oscillations can be seen both in the velocity map and the quasi-monochromatic map of normalized intensity.  Near the outer boundary of the superpenumbra, weak signs of oppositely directed flows, as one might expect from drainage of gas from fibrils, are apparent.  However, we stress that this structure must be confirmed with spectral interpretation since this map results simply from locating the spectral position of greatest \ion{He}{1} absorption and does not account for gradients along the line-of-sight.  

Absorption depth of the \ion{He}{1} triplet primarily correlates with the \ion{He}{1} number density and thickness of the absorbing regime, as well as with the degree of coronal illumination \citep{avrett1994}.  Photoionization of parahelium atoms and subsequent recombination is attributed to be the main driver for populating the orthohelium ground state \citep{centeno2008}.  While this study cannot constrain the relative degree of coronal illumination for each fibril, the visual correspondence of the \ion{He}{1} absorption with the H$\alpha$ fibrils is at least consistent with fibrils being regions of increased mass density as suggested by \cite{leenaarts2012}.


\section{\ion{He}{1} Triplet Polarization Signatures Within the Superpenumbral Region}\label{sec:spectra_description}

Anisotropic radiative pumping (i.e. the quantum extension to classical scattering), the Hanle effect, and the Zeeman effect work together to induce and modify the polarization of the \ion{He}{1} triplet.  To describe and heuristically interpret the polarization signatures observed here, we summarize some key aspects of these mechanisms. See \refasen, \cite{trujillo_bueno_2007}, and references therein for a more complete discussion.

\begin{deluxetable}{cccccl} 
\tablecolumns{4} 
\tablecaption{\ion{He}{1} Triplet Spectral Lines} 
\tablehead{\colhead{Transition} & 
	\colhead{Air Wavelength \mbox{\AA}}  &
	\colhead{$J_{lower}$} & 
	\colhead{$J_{upper}$}}
\startdata
$2s^{3}S_{1} - 2p^{3}P_{0}$ & 10829.0911 & 1 & 0 \\
$2s^{3}S_{1} - 2p^{3}P_{1}$ & 10830.2501 & 1 & 1 \\
$2s^{3}S_{1} - 2p^{3}P_{2}$ & 10830.3398 & 1 & 2
\enddata 
\label{tbl:triplet_lines}
\end{deluxetable} 

The \ion{He}{1} triplet consists of the three spectral lines (Table~\ref{tbl:triplet_lines}) formed between the $2p^{3}P$ and the $2s^{3}S$ terms of the \ion{He}{1} atom.  The two longer wavelength transitions form a blended `red' component at solar temperatures, whereas the 10829.0911 \mbox{\AA} transition is referred to as the `blue' component.  For each line, the Zeeman effect works as normal to split the degenerate energy transitions into $\pi$ ($\Delta M = 0$) and $\sigma$ ($\Delta M = \pm1$) components, with their respective polarizations.  Polarization is also induced (and modified via the Hanle effect) for each line by any ``order'' present in the magnetic substates of a given level participating in the transition.  This ``order'', what is termed atomic-level polarization, can be generated by ``order'' (i.e. anisotropy) in the incident radiation interacting with the atomic system.  The decay of an excited level or absorption from a lower level harboring ordered substates \textit{selectively} determines the polarization of the outgoing radiation.  Note, however, that unlike the other levels of the \ion{He}{1} triplet, the upper J-level (i.e. the total angular momentum) of the blue component is zero.  This makes it ``unpolarizable'' in the sense that there are no magnetic substates for which imbalanced populations and/or coherences can be produced via anisotropic radiative pumping.  Only \textit{selective absorption} processes create net polarization in the blue transition, while \textit{selective emission} and \textit{selective absorption} both contribute to the total polarization of the red component.  Multiterm calculations as in \refasen{ }are necessary to detemine the atomic-level polarization for each level since the ``order'' of a particular J-level can be transferred to other levels.  In fact, this process, called \textit{repopulation pumping} is important in determining the polarization of the lower-level of the \ion{He}{1} triplet.  The short-lived excited states of the $2p^{3}P$ term often map their ``order'' onto the meta-stable lower level.  

\begin{figure}
\epsscale{0.5}  
\plotone{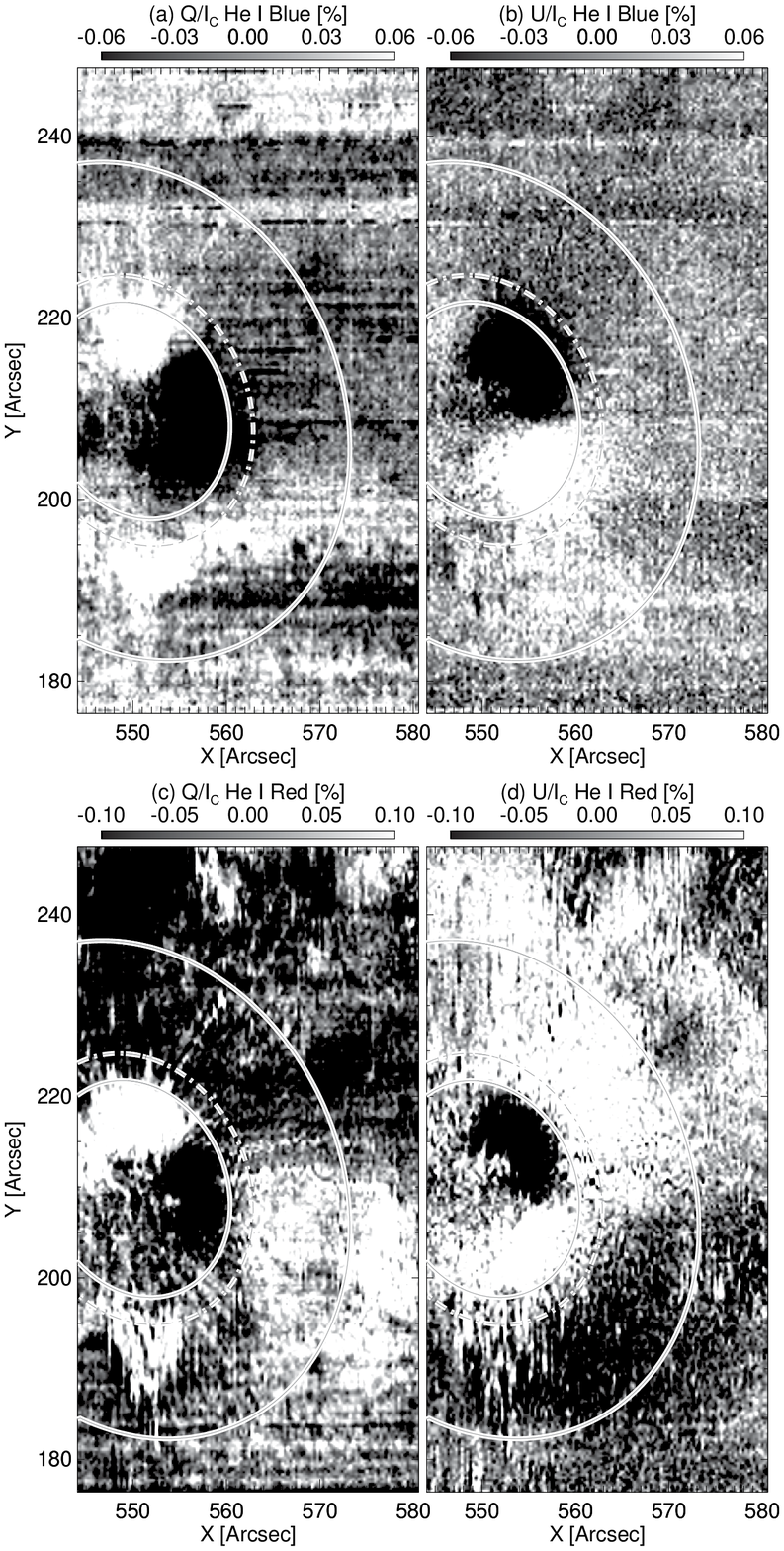}
\caption{Polarized maps of NOAA 11408 in the \ion{He}{1} triplet at 10830 \mbox{\AA}.  The top figures give Stokes Q and U for the \ion{He}{1} blue component while the bottom plots show the \ion{He}{1} red component.  Solar north is up and the reference direction for Stokes Q is solar east/west.  The color table is saturated as the levels given in the color tables to show the polarization of the superpenumbral region.  Horizontal streaks in the blue component maps arise from residual small-amplitude systematic measurement errors.  The solid lines show the boundaries of the sunspot penumbra ($r = r_{spot}$) and superpenumbra ($r = 2.3r_{spot}$).  The dashed lines corresponds to $r = 1.25r_{spot}$.  \label{fig:map_stokes}}
\end{figure}

Considering the distinctions between the blue and red \ion{He}{1} components, we compare and describe linearly polarized maps of the observed region in each component (see Figure~\ref{fig:map_stokes}).  To facilitate this description, we define a cylindrical geometry centered on the sunspot with a reference axis vertical in the solar atmosphere and a reference plane tangent to the solar surface.  The reference direction of the polar axis points towards disk center in correspondence with the scattering event geometry defined in \refasen.  Thus, the ray lying in the reference plane pointing from sunspot center towards disk center defines where $\theta_{spot}$ is equal to zero (see Figure~\ref{fig:fov_maps}).  The center of the sunspot is defined via a fit of a circle to the \textit{heliographic} coordinates of the external boundary of the photospheric penumbra on the north and east sides of the spot where the penumbra is most homogeneous in its radial extension.  This circle defines the sunspot radius (i.e. $r_{spot}$).  The superpenumbral fibrils extend mostly radially outward and have apparent endpoints within approximately 2.3 times the sunspot radius as shown in Figure~\ref{fig:fov_maps}.  A second set of fibril then extend in the same general direction from near this boundary (see Figure~\ref{fig:fov_maps}).

\begin{figure*}
\epsscale{1.}
\plotone{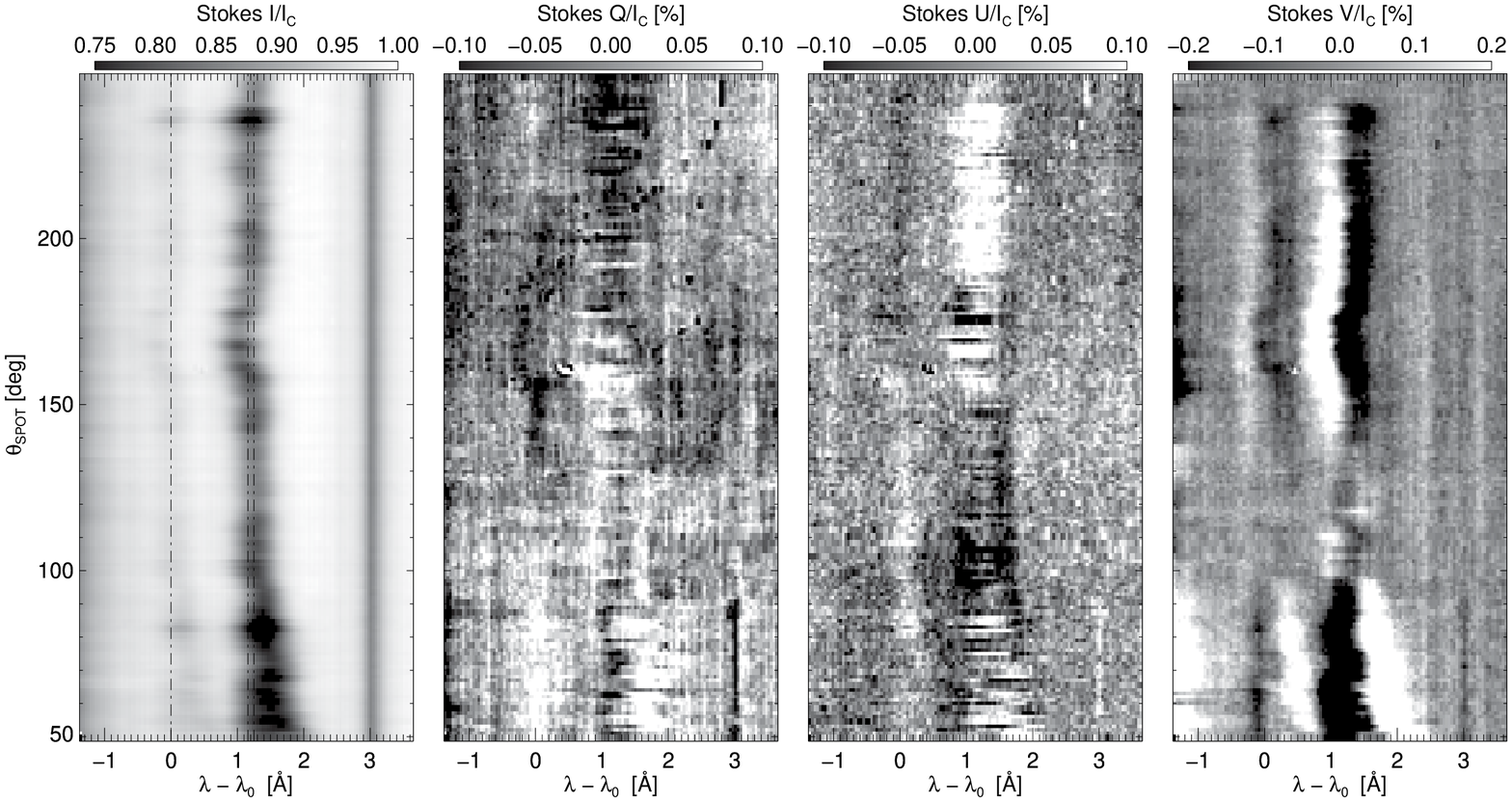}
\caption{Angular variation of the \ion{He}{1} Stokes profiles around the sunspot at a constant radius of 1.25 times the sunspot radius.  The vertical dashed lines indicate the rest wavelengths of the three \ion{He}{1} triplet transitions.  Only weak traces of the transverse Zeeman effect are noticeable in the linearly polarized Q and U profiles.  Rather, the mostly single-peaked, nearly Gaussian, characteristics of the Q and U profiles along with the opposite signed polarization in the blue and red components offer clear evidence for the prevalence of scattering polarization induced in the superpenumbra. \label{fig:ang_stokes}}
\end{figure*}

\subsection{The Zeeman-Effect Dominated Region}

Within the sunspot radius, the Q and U polarized maps for the blue and red components of the \ion{He}{1} triplet display the familiar lobe structure that is commonly seen in such maps of sunspots within photospheric normal Zeeman triplets.  The \ion{He}{1} component profiles in this region (not shown) exhibit a near symmetry about line center with the three nodes typical of the normal transverse Zeeman effect.  In the photosphere the four-lobe macroscopic structure found in Stokes Q and U maps centered on a nearly circular sunspot and viewed not too far from disk center ($\mu > 0.6$) is naturally explained by a simple model of a unipolar sunspot whose field diverges symmetrically as a function of height.  The four lobes appear as a consequence of the change in the azimuthal angle of the transverse component of the magnetic field relative to the line of sight (see, e.g., \cite{schlichenmaier2002}).  So too such a model is consistent with the global Q and U behavior exhibited within the sunspot radius by the \ion{He}{1} components in Figure~\ref{fig:map_stokes}.  As these maps are produced in the line center of each component, they exhibit the polarization of the on-average unshifted $\pi$-component.  Assuming the radiation absorbed originates as the unpolarized photospheric continuum, the linear polarization of the $\pi$-component induced by the transverse Zeeman effect is aligned perpendicular to the transverse component of the magnetic field relative to the line-of-sight.  With the simple model sunspot in mind, one would expect Stokes U to be near zero and Stokes Q to be negative along the observational reference direction for Stokes Q in a Zeeman-dominated region.  This is consistent with the pattern shown inside the sunspot radius in Figure~\ref{fig:map_stokes} as here the reference direction for Q is in the solar east/west direction (i.e. left/right in the figure).  This region can be considered Zeeman-dominated, and consequently the magnetic field strengths are expected to be on order of a kilogauss. 

\subsection{Atomic-Level Polarization in the Superpenumbral Region}

Just beyond the external boundary of the sunspot penumbra (i.e. $r = r_{spot}$), the global pattern of the linear polarization markedly changes particularly in the red component.  The sign of Q and U in the red component changes sign with respect to the lobe structure in the Zeeman-dominated region.  This is a consequence of the atomic level polarization beginning to dominate the polarization of the transverse Zeeman-effect.  To better illustrate this, we extract the observed profiles at a constant distance from sunspot center ($r = 1.25{ }r_{spot}$) and display them in Figure~\ref{fig:ang_stokes} stacked according to their angular position where $\theta_{spot} = 0$ points towards disk center.  For $\theta_{spot}$ greater than $90^{\circ}$ and less than $250^{\circ}$, Stokes Q and U show profiles clearly influenced by atomic-level polarization which can be distinguished despite the high level of noise in Stokes Q.  Only traces of a Zeeman-induced $\pi$-component situated between $\sigma$ components of opposite sign are recognizable in the \ion{He}{1} red component for $140^{\circ} > \theta_{spot} > 90^{\circ}$.  Furthermore, the red and blue components exhibit opposite signs indicating that selective absorption and emission processes are at work as discussed above.  From an observer's point of view, the opposite signs of the two components gives a beneficial indication that the level of intensity crosstalk in Stokes Q and U is negligible. 

Since Q and U are atomic-level polarization dominated in this region, the upper bound for magnetic field strengths is in the range of a few hundred Gauss ($B \lesssim 500$G).  The amplitude of Stokes V sets the lower bound, indicating field strengths are in the saturated regime of the Hanle effect.  For the \ion{He}{1} 10830 \mbox{\AA} triplet, the onset of Hanle saturation is near 8 G and 1 G for the upper level and lower level Hanle effect, respectively (\refasen).  Consequently, the degree of linear polarization is not sensitive to the magnetic field strength within the superpenumbral region, but is only sensitive to the direction of the magnetic field.  To constrain the full magnetic field vector of superpenumbral fibrils, we require the detection of a significant level of circular polarization (i.e. Stokes V).  For this reason, an oblique observing geometry where the sunspot is not too near disk center is preferred over a disk center perspective since fibrils are likely to be horizontal to the solar surface.

In the saturated Hanle regime, the population imbalances induced by anisotropic radiative pumping lead to linear polarization oriented either parallel or perpendicular to the horizontal component of the magnetic field \citep{trujillo_bueno_2007}.  Classical determinations of the scattering phase matrix (see, e.g., section 5.8 of \cite{landi_2004}) can be used to illustrate this principle for the classical analogue of an ``unpolarized'' lower level (i.e. $J_{l} = 0$) and a polarizable upper level with $J_{u} = 1$, and further illuminates the so-called Van Vleck effect.  Upper-level population pumping favors either the $\pi$ or the $\sigma{ }(\Delta M = \pm 1)$ transitions for this transition according to the Van Vleck angle $\theta_{V}$ defined when the angle between the radiation symmetry axis and the magnetic field is $54.74^{\circ}$.  For the classical analogue, one would expect selective emission processes to generate linear polarization parallel to the magnetic field for inclinations $\theta_{B}$ (w.r.t to the solar vertical and the radiation symmetry axis) greater than $\theta_{V}$ and less than $(\pi - \theta_{V})$, and linear polarization perpendicular to the magnetic field otherwise.  Multiterm calculations for the \ion{He}{1} red component indicate these expectations also apply to the collective behavior of the two red transitions.  Consider once again the simple model sunspot and make the common assumption that superpenumbral fibrils are oriented mostly horizontal to the solar surface(i.e. $\theta_{B} = 90^{\circ}$)  and extend radially from the sunspot. The induced atomic-level polarization for the red transitions is then expected to be parallel to the magnetic field, in fact the opposite of the polarization of the $\pi$-component for the Zeeman-dominated profiles discussed in the previous section.  The sign change in Q and U for the red component as a function of radius observed in Figure~\ref{fig:map_stokes} can thus be explained by and offers evidence for the near horizontal and radial orientation of superpenumbral fibrils.  

Lower-level \textit{depopulation} and \textit{repopulation} processes complicate the interpretation of the polarization of the blue \ion{He}{1} component (see \cite{trujillo_bueno_2002} for a complete discussion of optical pumping within the \ion{He}{1} triplet).  As can be seen in Figures~\ref{fig:map_stokes} and \ref{fig:ang_stokes}, the observed linear polarization in the blue component has the opposite orientation compared to the red component and can be seen to agree in pattern with the lobe-structure of the Zeeman-dominated region.  If one assumes that the decay of the polarized, short-lived upper levels accounts for the polarization of the lower-level in a \textit{repopulation} scenario, the linear polarization of the blue component should be opposite to the red component as observed in the superpenumbral region.  For the \ion{He}{1} triplet, this assumption can be validated for inclination angles away from the Van Vleck angle using the realistic multiterm calculations of \refasen.


\section{Analysis Methods}\label{sec:analysis}\label{sec:methods}

While the observed \ion{He}{1} polarization signatures offer heuristic support for the horizontal and mostly radial orientation of fibril magnetic fields, ultimately we wish to infer through inversion the magnetic field vector within individual superpenumbral fibrils, first to compare it with the visible fibril morphology, and then to study the three-dimensional magnetic architecture of the fibrils.  This requires first an efficient approach to select and model the visual aspects of each fibril.  Here we describe our approach as well as give details of the forward model and inversion method used to infer the magnetic field parameters from the fibril spectra.

\subsection{Fibril Tracing}

\begin{figure*}
\epsscale{1.}
\plotone{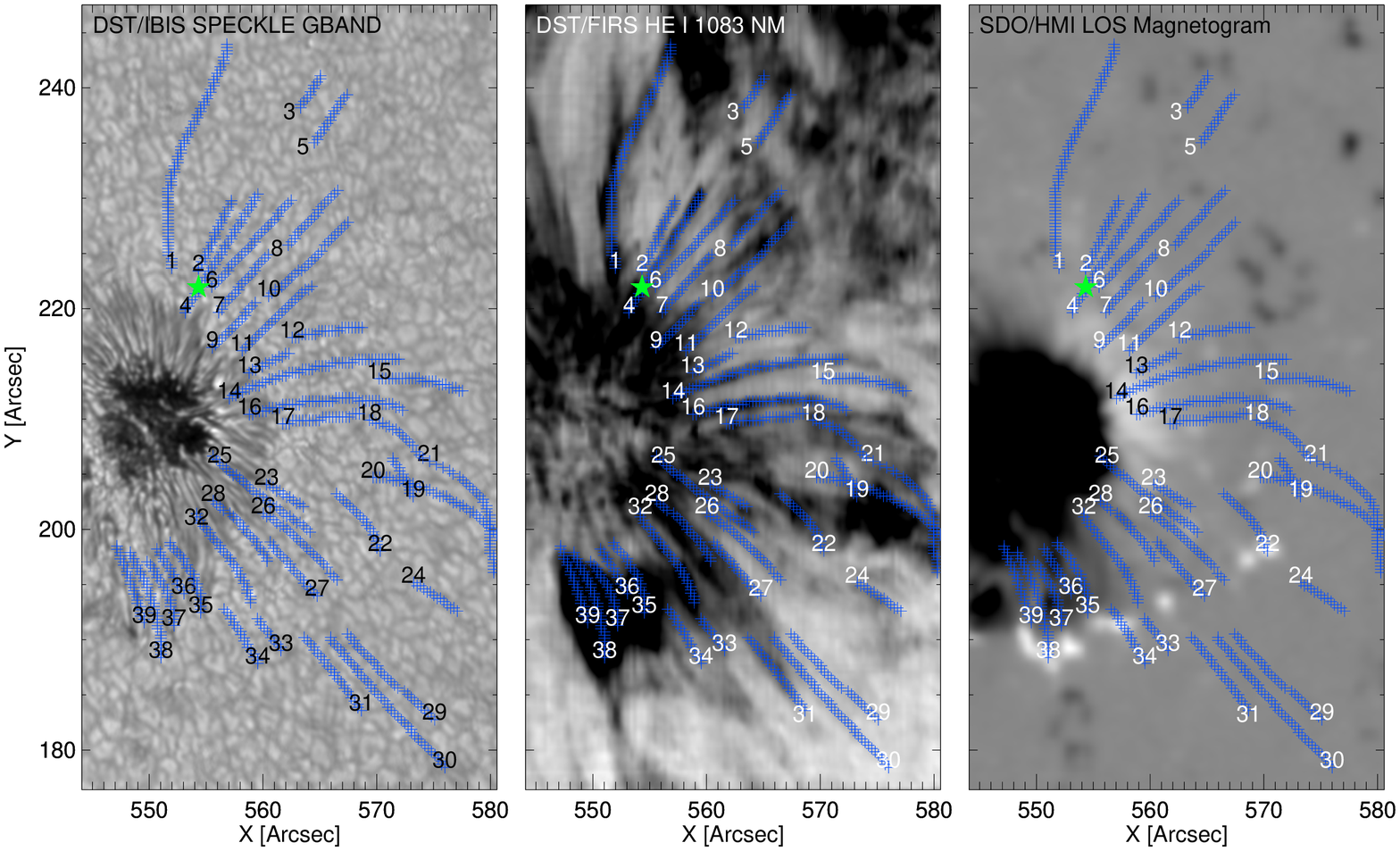}
\caption{Thirty-nine (39) fibrils manually traced by inspection of the entire observed \ion{He}{1} spectral data cube using CRISPEX.  Fibrils traced constitute curvilinear features of greater absorption and account for lateral variations of the Doppler shift and width of the \ion{He}{1} triplet.  The full Stokes spectrum of the point indicated by the yellow star is given in Figure~\ref{fig:ambig_spec}. (Color version available online) \label{fig:selected_fibrils}}
\end{figure*}

In the case of high-resolution narrowband imaging, \cite{jing2011} has suggested an automated way of selecting and modeling individual fibrils.  Using a threshold-based method, individual fibrils are located and fit with second-order polynomials.  The orientation angle of the fibril projected in the plane-of-the-sky (POS) is derived from the slope along this modeled curve.  Unlike single-channel narrowband imaging, the FIRS data contains the full spectral information of the \ion{He}{1} triplet over the full FOV.  It is advantageous to capitalize on the additional information as the spectral features of individual fibrils often exhibit variations in their Doppler velocity and Doppler width along their axis.  Narrowband imaging alone may restrict the full characterization of a fibril due to these variations.

The locations of individual fibrils in our FIRS data are found using the CRisp SPectral EXplorer \citep[CRISPEX:][]{vissers2012}, which is a widget-based visualization tool capable of quickly exploring spectroscopic and spectropolarimetric data cubes.  A particularly useful feature of CRISPEX allows the selection of points along loop-like features.  Since we can explore the spectral direction of the data cube while selecting points, we can trace fibrils exhibiting variations in Doppler velocity and/or width along their length.  In this way, we locate 39 fibril features in our FIRS scan (see Figure~\ref{fig:selected_fibrils}).

Once the points of the individual fibrils are selected, we model their projected morphology to derive their orientation angle in the POS.  As in \cite{jing2011}, we model each fibril with a simple functional fit (i.e. $y = f(x)$), defined in our case by an n-th order polynomial.  Most fibrils except, for example, superpenumbral whorls can be fit in this way.  Sometimes we must rotate reference axes to ensure that the selected fibril locations are fit as a function of the direction along its primary axis.  The derived orientation angles, which we also refer to as visible or ``traced'' orientation angles, can then be transformed into a common geometry.

Each fibril is fit to an appropriately ordered polynomial selected according to the Bayesian Information Criterion \citep{schwarz_1978, asensio_ramos_2012}:
\begin{equation}
{BIC = \chi^{2}_{min} + k \ln N},
\end{equation}
where $\chi^{2}_{min}$ is the normal summed squared difference between the data points and the fitted model, $N$ is the number of data points, and $k$ is the number of free parameters in the model (i.e. a n-th order polynomial has $k=n+1$ free parameters).  The best model is selected as the one which minimizes the BIC.  Of the 39 traced fibrils, 16 are best fit by a 1st order linear function, 14 by a 2nd-order, 8 by a 3rd-order, and 1 by a fourth-order polynomial. We collect the spectra along the modeled fibril axis at points found from a parametric cubic spline interpolation with a interval distance of $0.3''$.  In total, 985 individual fibril Stokes spectra are selected for analysis.  The projected orientation angle for each sampled location along the fibril is calculated from the first derivative of the fitted polynomial model, which for a curvilinear feature in the POS includes an intrinsic $180^{\circ}$ ambiguity. 

\subsection{Inversions of the \ion{He}{1} Triplet}\label{sec:hazel_description}

HAZEL refers to the advanced ``Hanle and Zeeman Light" forward modeling and inversion tool developed by \refasen.  Based on multiterm calculations \citep[see][]{landi_2004} of a five term model of the orthohelium atomic system, HAZEL determines the population imbalances and quantum coherences induced by anisotropic radiative pumping using the framework of the atomic density matrix.  The absorption and emission coefficients follow from the elements of the density matrix calculated via the statistical equilibrium equations subject to a limb darkened cylindrically symmetric radiation field whose symmetry axis is the solar vertical.\footnote{We ignore the symmetry-breaking effects of the sunspot.  The degree of symmetry-breaking due to a spot can be expressed as a function of its solid angle as viewed from a given point in the atmosphere (see section 12.4 of \cite{landi_2004}).  A sunspot of small solid-angle as viewed from the fibrils introduces a weak symmetry-breaking of the impringing radiation field that weakly influences the emergent polarization of the fibrils.  We are unable to discern the role of this symmetry-breaking from these measurements. While the sensitivity shown here is great, we require a further reduction of the noise to evaluate this effect.  We leave this for future work. }  HAZEL also correctly accounts for the Hanle, Zeeman, and Paschen-Back effects.  See \refasen{ }for more details.  

We use the inversion capability of HAZEL to interpret the 985 observed \ion{He}{1} Stokes spectra from the 39 selected fibrils.  The equation defining the radiative transfer for each Stokes vector is an exact analytical solution of a constant-property slab model including magneto-optical terms and stimulated emission.  The slab is described by the following deterministic quantities: thermal Doppler broadening $v_{th}$, macroscopic line-of-sight velocity $v_{mac}$, optical thickness $\Delta\tau$, damping parameter $a$, magnetic field $B$, magnetic field inclination angle $\theta_{B}$, and magnetic field azimuthal angle $\chi_{B}$ at a constant height $h$.

\begin{figure}
\epsscale{1.}
\plotone{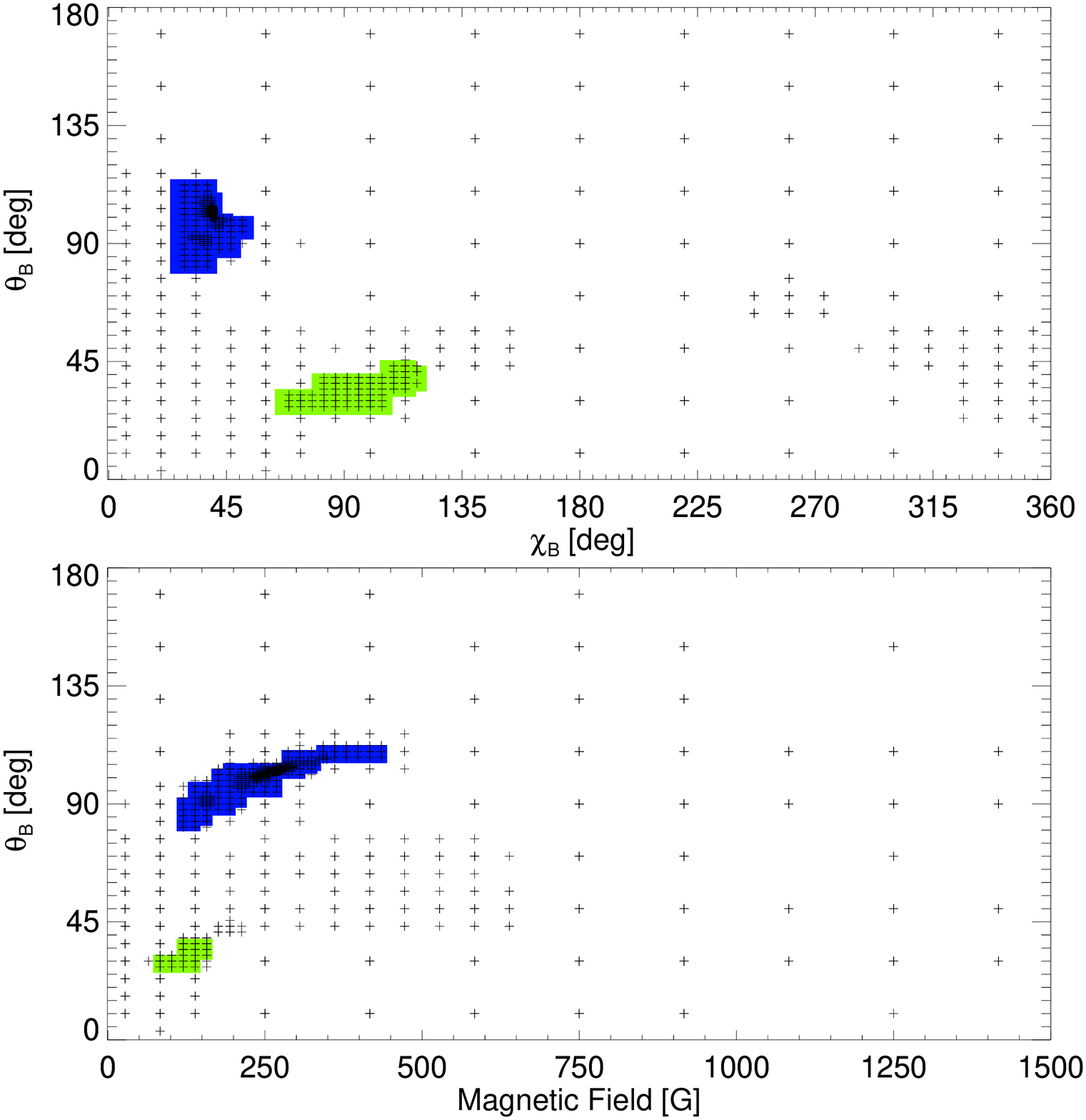}
\caption{Illustration of the hyper-volume search performed by the DIRECT algorithm used to locate solution ambiguities, as in \cite{asensio_ramos2008}.  Each data point represents one point (of 3000) in the hyper-volume for which DIRECT calculates the $\chi^{2}$ parameter, as defined in Equation~\ref{eq:chi_total}.  The colored regions are areas where $\chi^{2} < \chi^{2}_{min} + 0.25$, our prescription for regions of probable ambiguities. (Color version available online)  \label{fig:direct_results}}
\end{figure}

The oblique geometry of the observed region greatly influences the manner in which we use HAZEL.  As described by \refasen{ }and \cite{merenda2006}, \ion{He}{1} spectropolarimetric observations at disk center and off-limb are subject to two ambiguities:  1) the Van Vleck ambiguity for some range of inclinations and 2) the familiar $180^{\circ}$ ambiguity which introduces a $180^{\circ}$ azimuth ambiguity at disk center and off-limb, and in addition the off-limb Stokes spectra for $\left \langle B,\theta_{B},\chi_B \right \rangle$ are $\left \langle B,180^{\circ} - \theta_{B},-\chi_B \right \rangle$ are indistinguishable.  According to \refasen{ }and \cite{trujillo_bueno_2010}, an oblique scattering angle introduces a quasi-degeneracy associated with having the preferential axis of the Zeeman effect and that of the radiation symmetry be different from $0^{\circ}$ or $90^{\circ}$.  This degeneracy is lifted for large oblique angles such that the Zeeman effect and atomic-level polarization work to remove some ambiguities \citep[see][]{landi1993}; though what ambiguities remain can be difficult to determine.  Moreover, observational noise can introduce ambiguities.  We thus rely on the DIRECT algorithm \citep{jones1993} described in \refasen{ }to search for ambiguous solutions for all possible field geometries (i.e. $0^{\circ}<\theta_{B}<180^{\circ}$,$0^{\circ}<\chi_{B}<360^{\circ}$).

Our inversion scheme relies on the available HAZEL tools with a different implementation than \refasen.  We find the thermodynamic parameters should not be fit independently with fits only to Stokes I.  Fibril spectra near the sunspot exhibit significant Zeeman magnetic broadening.  We do not single out these spectra.  Rather, we create a standardized approach for all spectra.  First, we remind the reader that optimizing the model solution involves minimizing the reduced chi-squared merit function (\refasen):
\begin{equation}
{\chi^{2} = \frac{1}{4}\sum_{i=1}^{4} w_{i}\chi^{2}_{i}}.\label{eq:chi_total}
\end{equation}
where the individual contributions of the chi-squared to for each Stokes parameter is
\begin{equation}
{\chi^{2}_{i} = \frac{1}{N_{\lambda}}\sum_{j=1}^{N_{\lambda}}\frac{[S_{i}^{syn}(\lambda_{j}) - S_{i}^{obs}(\lambda_{j})]^2}{\sigma_{i}^{2}(\lambda_{j})}}\label{eq:chi_individual}
\end{equation}
$\chi^{2}_{i=0,1,2,3} =\chi^{2}_{I,Q,U,V}$ and all other variables are as defined in \refasen.  We choose a 4-step approach similar to \refasen.  First, the thermodynamic parameters are found via the DIRECT algorithm with the model parameters $v_{th}$, $v_{mac}$, $\Delta\tau$,  $B$, $\theta_{B}$, and $\chi_{B}$ free to vary within a realistic range.  Such a great number of free parameters reduces the convergence efficiency of the DIRECT approach due to an increased dimensionality which must be compensated for by a greater number of function evaluations.  We improve the efficiency by weighting the Stokes I chi-squared (i.e. $\chi^{2}_{I}$) greater than the others (i.e., $w_{I} = 5, w_{Q,U,V} = 1$).    Furthermore, the damping parameter, $a$ is not a free parameter.  We calculate $a$ directly from the Doppler width, $v_{th}$, using the Einstein coefficient of the transition, which accounts for thermal effects only.  In a second step, the free parameters are fine-tuned using the Levenburg-Marquardt (LM) method of HAZEL.  We find this procedure gives much more reliable determinations of the thermodynamics parameters for all spectra.

\begin{figure*}
\epsscale{1.}
\plotone{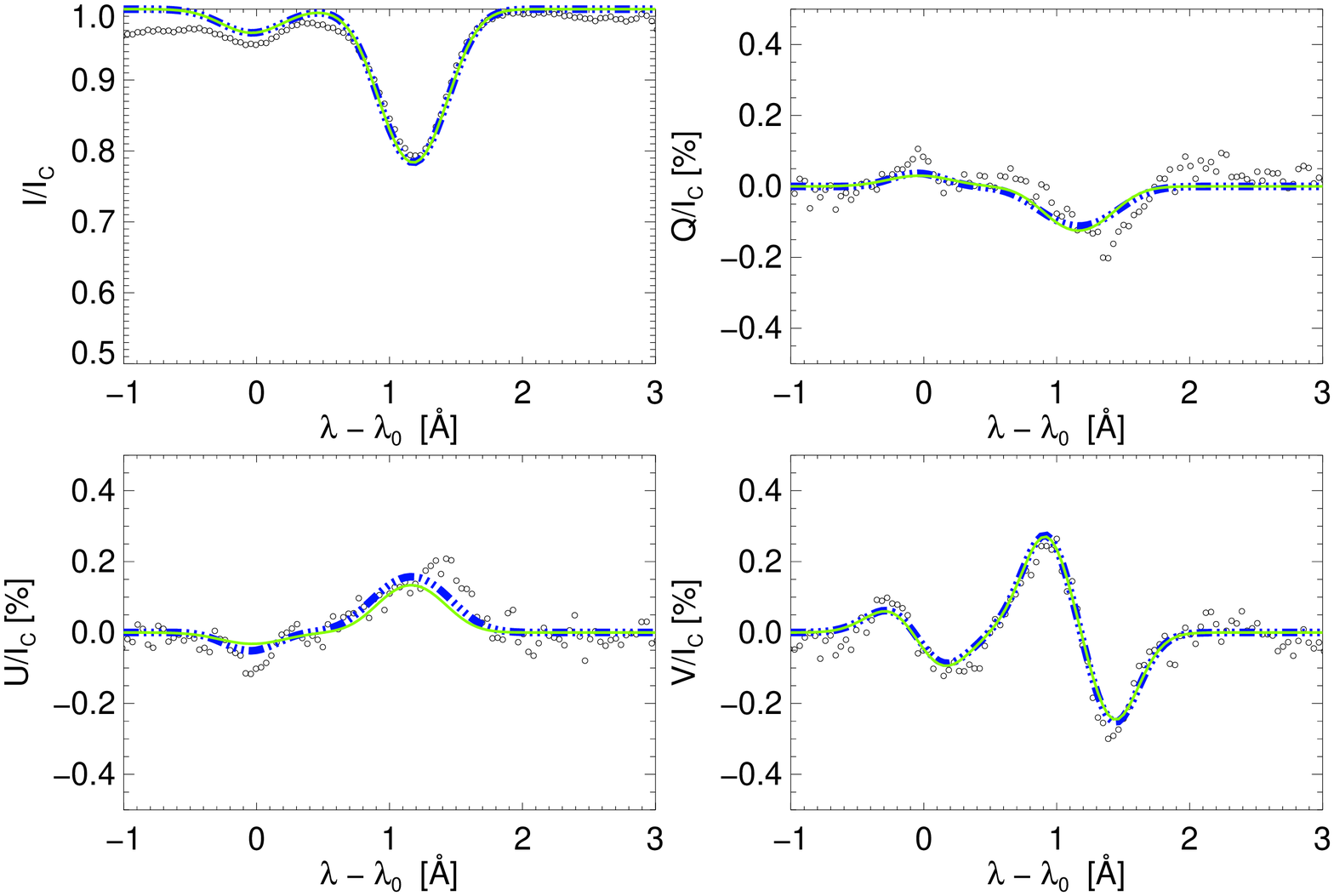}
\caption{Observed \ion{He}{1} triplet Stokes spectrum taken from a superpenumbral fibril (open circles) and fit with HAZEL inversions.  Two solutions are found with similarly good fits and correspond to the two regions identified in Figure~\ref{fig:direct_results}.  The blue solid line corresponds to the best fit magnetic field vector of $B = 262$ G, $\theta_{B} = 101^{\circ}$, $\chi_{B} = 40^{\circ}$, with $\chi^{2}_{Q,U,V} =  \{ 1.23847,1.12328,1.51492\}$.  The yellow dot-dashed line corresponds to the slightly poorer fit using $B = 139$ G, $\theta_{B} = 34^{\circ}$, $\chi_{B} = 100^{\circ} $, which yields $\chi^{2}_{Q,U,V} =  \{ 1.26013, 1.62421, 1.50613 \}$. (Color version available online) \label{fig:ambig_spec}}
\end{figure*}

Once the thermodynamics parameters ($v_{th}$, $v_{mac}$, and $\Delta\tau$) are determined, we locate all relevant solutions for the magnetic field strength and direction.  We use the method suggested by \refasen{ }that exploits the properties of the deterministic DIRECT searching algorithm, a key component of which is that no region of the parameter space is entirely eliminated from the search process over a great number of iterations.  Ambiguous solutions are found via systematic searching of the parameter space.  The number of function evaluations increases in regions of the parameter space resulting in better model fits, forming clusters in maps of the searching process (see Figure~\ref{fig:direct_results} here, and Figure 17 of \refasen).  We allow for a total of 3000 function evaluations by the DIRECT algorithm to locate these regions in the $B$, $\theta_{B}$,$\chi_{B}$ hyper-volume (i.e. n-dimensional parameter space), during which all Stokes parameters are weighted equally in the merit function (Equation~\ref{eq:chi_total}).  Up to three regions are then identified in the DIRECT searching maps according to two criteria: 1) the possible solutions must be separated in the $\theta_{B}$,$\chi_{B}$ space (top of Figure~\ref{fig:direct_results}); and 2) the best fit in the identified cluster must be less than $\chi^{2}_{min} + \delta_{\chi}$, where $\chi^{2}_{min}$ is the minimum reduced chi-squared found by the DIRECT algorithm.  We choose a value of $\delta_{\chi}$ equal to $0.25$, as this is the change in $\chi^{2}_{min}$ induced by a 1$\sigma$ error in any one of $\chi^{2}_{I,Q,U,V}$.  The result of this process for the spectra from the location marked by a yellow star in Figure~\ref{fig:selected_fibrils} is illustrated in Figure~\ref{fig:direct_results}.  As seen in the figure, two  $\theta_{B}$,$\chi_{B}$ subspaces are identified as possible solutions.  The best fit in each subspace is identified and fine-tuned in a fourth inversion step which employs once again the LM method initialized with the identified DIRECT solutions.  These determined parameters are then our best fit determinations of the plasma thermodynamic and magnetic properties (see Figure~\ref{fig:ambig_spec}).

We do not fit the height of the plasma as a free parameter.  Population imbalances are a function of the anisotropic properties of the pumping radiation field.  Due to geometry, the degree of anisotropy varies as a function of height, as so too does the mean intensity; albeit, this dependence is weaker (\refasen)\footnote{A pitfall we discovered here is that this mean intensity change also effects the determination of $\Delta\tau$ via the emission coefficient $\bf{\epsilon_{I}}$. $\Delta\tau$ then is a weak function of height and influences the goodness-of-fit, meaning $\Delta\tau$ needs to be a free parameter investigated alongside the height dependence of the polarized spectra}. \cite{merenda2011} used this principle to infer the height of chromospheric material above an emerging flux region.  \refasen{ }noted this possibility but also described that a quasi-degeneracy between height and inclination can make it difficult to infer the height without an additional constraint on the field geometry.  We elect, due to the level of noise in our observations, to keep the height as an assumed constant and then discuss the influence of this choice in section~\ref{sec:infer_heights}.  The NLTE calculations of \cite{centeno2008} indicate a large range of heights contributing to \ion{He}{1} absorption.  For the 1D FAL atmospheric model, the range of \ion{He}{1} formation is between 1 and 2.2 Mm above the solar surface.  This is consistent with the correspondence between \ion{He}{1}, H$\alpha$, and \ion{Ca}{2} described in section~\ref{sec:multi_wave}.  The primary contribution to the \ion{He}{1} absorption is consequently thought to be within a fibril's depth of the contribution peak of the H$\alpha$ and \ion{Ca}{2}.  \cite{leenaarts2012} argues that fibrils in H$\alpha$ are formed at higher relative heights than the inter-fibril plasma and showed fibril formations ranging from 1.5 to 2.75 Mm.  We fix the height used for the inversions at 1.75 Mm in accordance with these observations.


\section{Results}\label{sec:results}

\begin{figure*}
\epsscale{1.}
\plotone{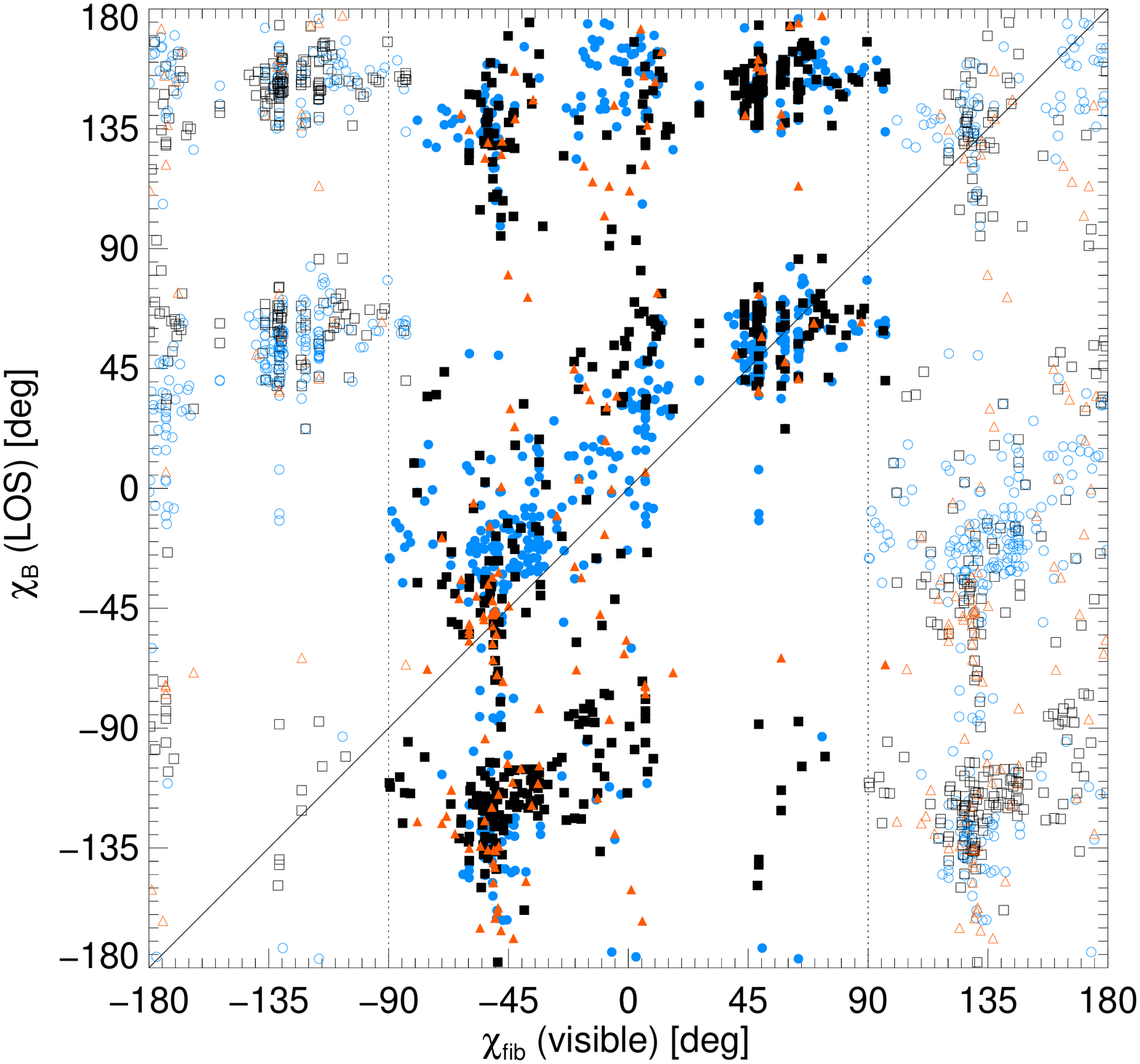}
\caption{Comparison of all determinations of the projected angle of the inferred magnetic field vector (i.e. the line-of-sight azimuthal angle, $\chi_{B}$) and the visible projected angle of the observed fibrils, $\chi_{fib}$.  For each inverted spectrum, all the inferred solutions found via the HAZEL inversions are plotted versus the two values of the visible projected azimuth, which has an inherent $180^{\circ}$ ambiguity.  Blue circles, black squares, and orange triangles, respectively give the first, second, and third best solution for any single HAZEL inversion.  Open and closed data points distinguish between the two ambiguous visible directions. (Color version available online)  \label{fig:azi_comp}}
\end{figure*}   

In Figure~\ref{fig:azi_comp} we plot the full results of the section~\ref{sec:methods} analysis for every selected fibril location at which the Stokes spectrum is reasonably well fit with the HAZEL inversion method (i.e. $\chi^{2}_{I,Q,U} < 2.5$).  A total of 592 (of 985) spectra meet this criteria.  The figure includes the effect of all ambiguities.  On the x-axis, $\chi_{fib}$ is the projected angle of the observed ``visible'' fibril as manually traced in the POS with the inherent $180^{\circ}$ ambiguity.  For each value of $\chi_{fib}$, the y-axis reports the azimuths, $\chi_{B}$, of the inferred magnetic field solutions resulting from the inversion of the observed Stokes spectra, and transformed into the line-of-sight geometry (i.e. the projected angle of the magnetic field transverse to the line-of-sight).  The reference direction for both $\chi_{fib}$ and $\chi_{B}$ points toward solar west.  Note the already strong correlation of many of the solutions in this plot which represents the affect of all ambiguities.  The ordered nature of this figure is due to the non-random influence of the Van Vleck and $180^{\circ}$ ambiguities.  When these effects are taken into account, as discussed below, this figure provides direct evidence for the field alignment of superpenumbral fibrils well outside of the penumbral boundary.

\subsection{Variation of the Magnetic Field Vector Along Individual Fibrils}\label{sec:individual_fibril}

\begin{figure*}
\epsscale{1.}
\plotone{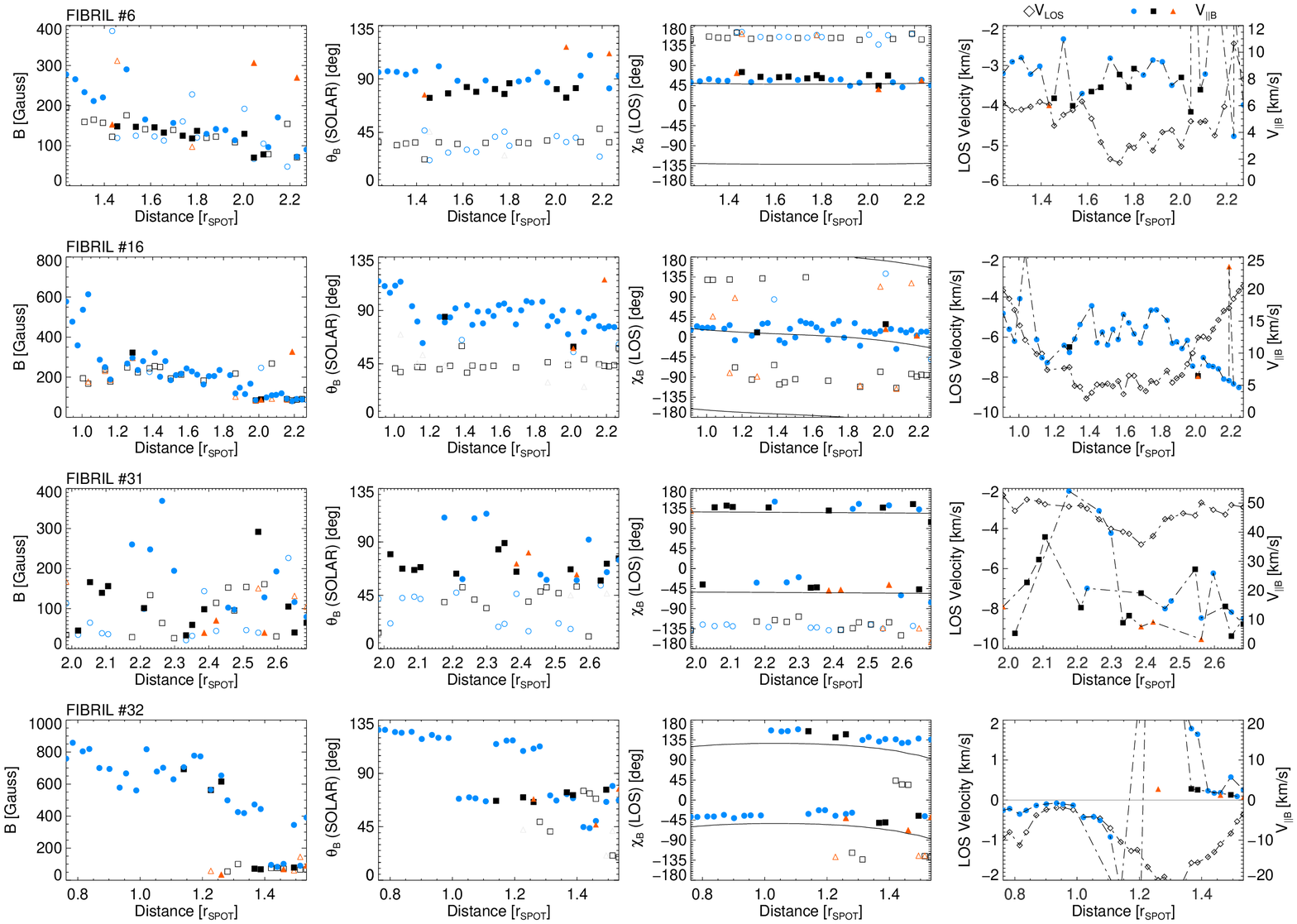}
\caption{Derived magnetic field parameters along fibrils 6, 16, 31, and 32 identified in Figure~\ref{fig:selected_fibrils}. All found solutions are represented, with blue circles denoting the best fit solution.  Filled data points correspond to the solutions matching the visible azimuth determined via tracing the fibril (solid lines). See text for details. $r_{spot}\cong 8.7$ Mm. (Color version available online) \label{fig:fib_para}}
\end{figure*}

We concentrate on four representative fibrils (\#6,16,31,32 in Figure~\ref{fig:selected_fibrils}) and plot the full results as a function of distance from the sunspot center in Figure~\ref{fig:fib_para}.  Two different reference systems are used here to report the inclination and azimuth of the magnetic field vector.  The inclination refers to the angle between the magnetic field direction and the local solar vertical and is directly determined by HAZEL. An outward-directed radial magnetic field has $\theta_{B}=0^{\circ}$ and a value of $90^{\circ}$ refers to a horizontal magnetic field.  Meanwhile, to allow direct comparison, the azimuths of the magnetic field vector and the modeled fibril orientation angle are given in the line-of-sight geometry since one cannot transform the projected angle of the traced fibril into solar coordinates.  Lastly, since there is good reason to believe that the flows follow the magnetic field in this ion-neutral coupled, high electric conductivity plasma \citep{judge2010}, we plot the magnitude of the velocity directed along the magnetic field vector according to:  
\begin{equation}
v_{\left |  \right | B} = \frac{-v_{LOS}}{\cos{\theta_{B,LOS}}}\label{eq:proj_vel}
\end{equation}
where $\theta_{B,LOS}$ is the inclination of the magnetic field in the line-of-sight geometry, $v_{LOS}$ is the velocity projected along the line-of-sight (negative values correspond to velocities of approach, i.e. blue-shifted spectra), and $v_{\left |  \right | B}$ is the velocity projected along the derived magnetic field.  With the negative sign in Equation~\ref{eq:proj_vel}, we assign negative values of $v_{\left |  \right | B}$ to flows that are anti-parallel to the magnetic field direction.

Let us first compare the two determinations of the azimuth in Figure~\ref{fig:fib_para}.  For each fibril, different classes of solutions are found for $\chi_{B,LOS}$.  In the case of fibrils \#6 and \#16, only one of these classes (represented by filled data points in the figure) matches only one determination of the traced orientation angle.  For fibrils \#31 and \#32, both ambiguous determinations of the projected orientation angle are matched by a class of HAZEL solutions along the fibril.  This behavior can be explained by comparing the magnitudes of the LOS component of the magnetic field in each case.  For all ambiguous solutions found for the same Stokes spectra, the LOS field magnitude must be nearly the same, but subject to the role of observational noise.  The LOS magnitude of $\vec{B}$ for fibrils 6,16,31, and 32 are on average 81,126,25, and 57 Gauss respectively.  If one compensates for the large field strengths of \#32 near the sunspot, a lower value would be more representative of the LOS field magnitude.  Fibrils 31 and 32 show fields oriented more perpendicular to the LOS than 6 and 16, with lower values of $\left | B_{LOS} \right |$. This results in Stokes V signals of lower amplitude, at or close to the level of the noise.  Consequently, the Stokes V spectra of fibrils 31 and 32 cannot be used to distinguish between the two azimuth solutions, whereas in fibrils 6 and 16 this ambiguity is resolved with Stokes V.  Fields that are oriented nearly perpendicular to the LOS are subject to the $180^{\circ}$ Hanle ambiguity.

Now consider the additional information of the inclination angle.  It should be noted that the sunspot umbra hosts a field of inward directed polarity (i.e. $\theta_{B} \approx 180^{\circ}$).  The azimuth matched solutions of fibrils 6 and 16 are characterized by horizontal fields, and in the case of fibril \#16, the inclination increases near the sunspot reflecting a downwards turn into the sunspot and consistent with the polarity of the umbra.  The inclinations of 31 and 32 are noisier, but the one solution found for \#32 within the sunspot radius, $r_{spot}$, is consistent with the polarity of the sunspot umbra.  Furthermore, the azimuth of this solution (filled circles), when compared to the one matching solution of \#6, is consistent with the variation in the fibril direction around the sunspot.  As there is only one solution for fibril 32 with the sunspot's radius, we have an unambiguous determination of the fibril magnetic field and it is directed along the fibril axis.  For each fibril, the classes of solutions which do not match any traced fibril direction are characterized by inclinations near the sunspot below the Van Vleck angle and are discontinuous with the polarity of the sunspot.  These solutions are consistent with the influence of the Van Vleck effect on the inversion process and are ruled unphysical since the magnetic field is expected to be continuous along the fibril.  The picture of fibrils rooted in the sunspot that become nearly horizontal away from the sunspot is thus supported by these measurements.

\begin{deluxetable}{cccccl}
\rotate
\tablecolumns{6} 
\tabletypesize{\tiny}
\tablecaption{Comparison of Visible and Inferred Azimuthal Direction of Fibrils} 
\tablehead{\colhead{Fibril No.} & 
	\colhead{$\bar{\chi}_{fib}$(visible)\tablenotemark{a}}  &
	\colhead{$\bar{\chi}_{B}$(inferred)\tablenotemark{b}}  &
	\colhead{$\Delta \bar{\chi}_{n} / \sigma_{n}$\tablenotemark{c}} &
	\colhead{$\left | \bar{B}_{LOS} \right | $(inferred)}  &
	\colhead{Comments}}
\startdata
     1  &    77.921,          &    62.590           &  -1.641         &   39.0  & One additional Van Vleck induced solution \\
     2  &    63.862,          &    54.857           &  -0.933         &   86.5  & One additional Van Vleck induced solution \\
     3  &    59.239,          &    74.640           &   2.057         &   25.0  & One additional Van Vleck induced solution \\
     4  &    58.852,          &    51.105           &  -0.607         &   79.9  & One additional Van Vleck induced solution \\
     5  &    57.214,          &    54.102           &  -0.263         &   21.8  & One additional Van Vleck induced solution \\
     6  &    48.909,          &    58.194           &   1.077         &   81.4  & One additional Van Vleck induced solution \\
     7  &    50.074,          &    58.482           &   1.161         &  109.5  & One additional Van Vleck induced solution \\
     8  &    48.968,          &    43.365           &  -0.258         &   28.0  & One additional Van Vleck induced solution \\
     9  &    46.801,          &    48.367           &   0.174         &  122.3  & One dominant matching solution \\
    10  &    44.002,          &    47.319           &   0.252         &   51.8  & One additional Van Vleck induced solution \\
    11  &    43.622,          &    53.649           &   1.014         &  113.3  & One additional Van Vleck induced solution \\
    12  &     9.004, 188.567  &    54.124, 155.680  &   2.803,-3.912  &   79.2  & Two inconsistent solutions \\
    13  &    26.623,          &    49.956           &   1.693         &  154.7  & One additional Van Vleck induced solution \\
    14  &     9.662,          &    29.578           &   0.854         &  110.9  & Noisy spread in two Van Vleck induced solutions \\
    15  &   122.277,          &   160.680           &   3.731         &   43.8  & Two additional Van Vleck induced solutions \\
    16  &     0.312,          &    10.120           &   0.460         &  126.1  & Noisy spread in two Van Vleck induced solutions \\
    17  &     6.394,          &    13.633           &   0.391         &  160.1  & Two additional Van Vleck induced solutions \\
    18  &   -40.317,          &    -5.052           &   2.831         &   66.2  & One additional Van Vleck induced solution \\
    19  &   -64.574, 112.216  &   -21.747, 140.917  &   1.785,13.528  &   69.4  & One additional Van Vleck induced solution \\
    20  &   -24.941, 162.636  &   -19.952, 137.908  &   0.367,-2.392  &   53.0  & One additional Van Vleck induced solution \\
    21  &   -63.104, -63.104  &   -12.864,-109.765  &   4.143,-6.463  &   25.6  & Two inconsistent solutions \\
    22  &   -50.687, 123.633  &   -23.109, 143.933  &   2.732, 7.529  &   54.7  & One additional Van Vleck induced solution \\
    23  &   -33.498, 146.502  &   -12.147, 153.163  &   1.436, 0.456  &   30.9  & One additional Van Vleck induced solution \\
    24  &   -35.495,          &   -17.558           &   1.660         &   34.3  & One additional Van Vleck induced solution \\
    25  &   -37.024, 141.248  &   -23.240, 160.169  &   1.355, 1.911  &   41.2  & One additional Van Vleck induced solution \\
    26  &   -41.884,          &   -29.319           &   2.585         &   33.7  & One additional Van Vleck induced solution \\
    27  &   -50.588, 129.491  &   -40.223, 140.274  &   0.382, 0.600  &   33.5  & One additional Van Vleck induced solution \\
    28  &   -50.297, 130.065  &   -34.746, 139.278  &   1.745, 0.542  &   91.0  & Two dominant solutions, minor Van Vleck influence \\
    29  &   -45.863, 134.719  &   -40.235, 129.986  &   0.133,-0.242  &   16.5  & One additional Van Vleck induced solution \\
    30  &   -49.705, 130.312  &   -51.324, 127.301  &  -0.088,-0.198  &   16.1  & One additional Van Vleck induced solution \\
    31  &   -54.316, 125.811  &   -43.265, 136.224  &   0.752, 0.962  &   24.9  & One additional Van Vleck induced solution \\
    32  &   -61.479, 116.690  &   -45.945, 142.609  &   0.545, 2.486  &   56.6  & Two additional Van Vleck induced solutions \\
    33  &   -53.871, 126.138  &   -32.228, 137.541  &   2.439, 1.177  &   21.9  & One additional Van Vleck induced solution \\
    34  &   -59.724, 120.274  &   -50.920, 128.125  &   1.151, 1.125  &   17.5  & One additional Van Vleck induced solution \\
    35  &   -54.248, 126.856  &   -41.159, 112.941  &   1.382,-0.392  &  306.0  & Two dominant solutions, minor Van Vleck influence
\enddata 
\tablenotetext{a}{The average visible (i.e. traced) azimuth direction of each fibril has a $180^{\circ}$ ambiguity}
\tablenotetext{b}{Average, disambiguated azimuths inferred via HAZEL inversion are given in the line-of-sight geometry with a reference direction consistent with the azimuths found via tracing and modeling the fibril morphology}
\tablenotetext{c}{Total azimuthal angle error estimated to be approximately $20^{\circ}$ per spatial pixel}
\label{tbl:vis_vs_infer}
\end{deluxetable} 

We characterize all 39 analyzed fibrils in the same manner as above and detail these results in Table~\ref{tbl:vis_vs_infer}. The average values of the fibril orientation angles ($\bar{\chi}_{fib}$) and the inferred LOS azimuth of the magnetic field ({$\bar{\chi}_{B}$) are given only for the solutions not classified as Van Vleck ambiguities.  The deviation ($\left | \Delta \bar{\chi}_{n}\right | $) between the average orientation angle and the average LOS azimuth is recorded as a factor of the standard deviation ($\sigma_{n}$) of the LOS magnetic field azimuths, which represent the dominate source of error.  Only for fibrils 15, 21, and 22 do the inferred azimuths of the magnetic field ({$\bar{\chi}_{B}$) deviate more than $3\sigma$ from the traced orientation angle  ($\bar{\chi}_{fib}$).  These fibrils correspond to short fibrils selected in a complex area of the observed region far from the sunspot.  We expect that improper selection or modeling explains their $>3\sigma$ deviation in azimuth rather than a real misalignment of the field and fibril.  This table also reiterates that fields with a direction primarily transverse to the LOS introduces a $180^{\circ}$ ambiguity in the azimuth.  This affects those fibrils on the southwest side of the sunspot (Nos. 19-35).  The fibrils on the northwest side (Nos 1 -18, expect 12) all have one single inverted solution for the magnetic field azimuth that matches the azimuthal direction of the traced fibrils.  The spectra of fibrils 36-39 are not well fit with a one component HAZEL model. 

\subsection{Correlation of Visible and Inferred Azimuths} \label{sec:azimuth_comparison} 

\begin{figure*}
\epsscale{1.}
\plotone{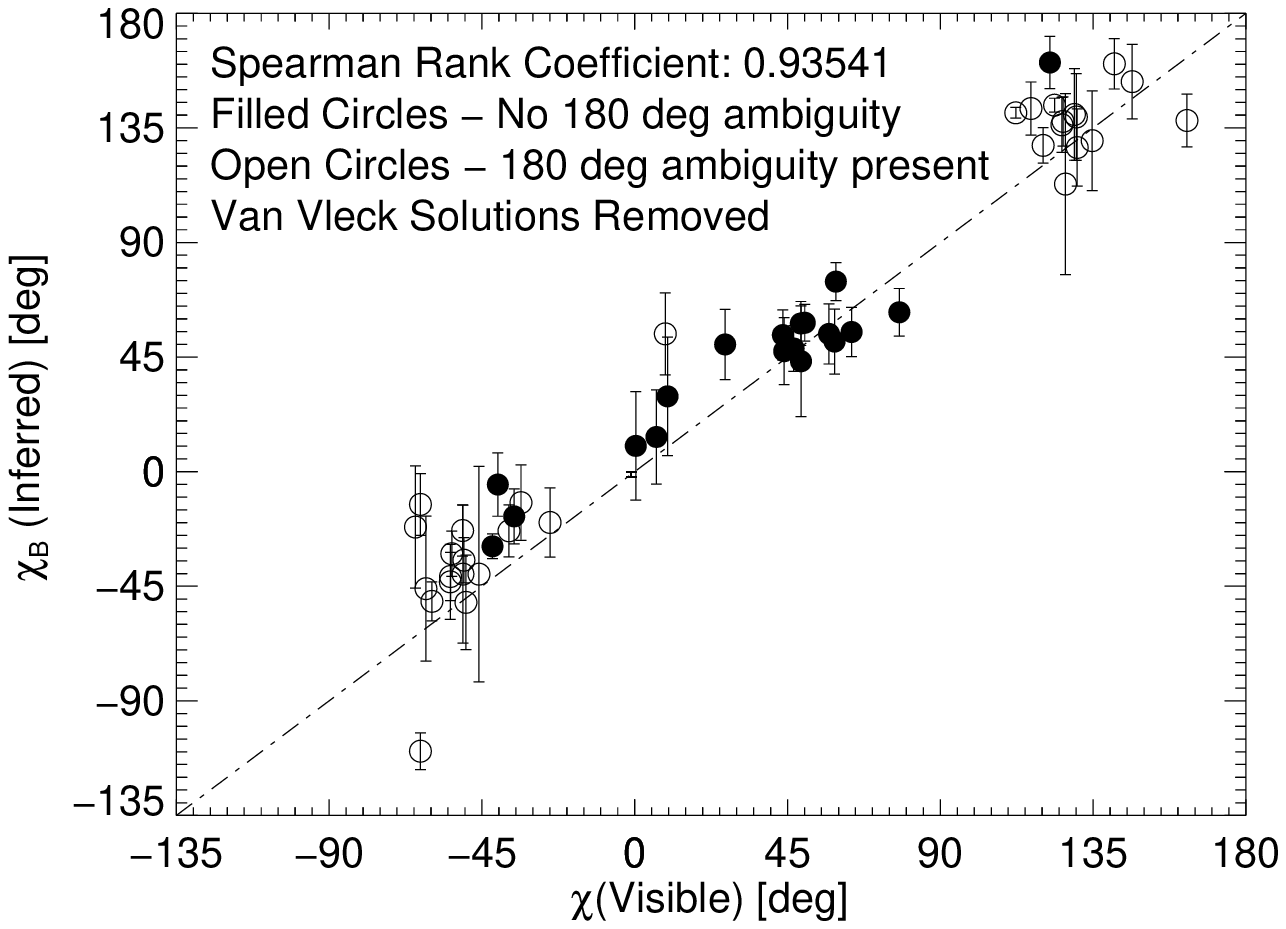}
\caption{Comparison of the average visible projected angle of the fibrils and the inferred, disambiguated average azimuthal direction of the magnetic field.  The visible azimuths have an inherent $180^{\circ}$ ambiguity and that all values are given in the line-of-sight geometry in which a vector with an azimuthal angle of zero points towards solar east.  Values are given along with the individual fibril numbers in Table~\ref{tbl:vis_vs_infer}. \label{fig:average_azi_comp}}
\end{figure*}   

Figure~\ref{fig:average_azi_comp} illustrates the results found in Table~\ref{tbl:vis_vs_infer} and is akin to Figure~\ref{fig:azi_comp} without the Van Vleck induced solutions.  Filled data points correspond to fibrils with only one matching azimuth, while open circles are plotted for the pair of solutions for the fibrils with a $180^{\circ}$ ambiguity in their azimuth.  These different situations are classified according to their azimuth angle.  A Spearman ranking test of these solutions give a correlation coefficient of 0.935.  We consider this the best proof to date that fibrils are visual markers for the magnetic field.

\subsection{Maps of Fibril Quantities} \label{sec:fibril_magnetic_maps} 

We investigate the spatial variation of the fibril magnetic field vector with spatial maps of the field parameters (see Figure~\ref{fig:fibril_field_maps}).  Although the parameter errors are not well known\footnote{Formal errors in an spectropolarimetric inversion is a matter of current research.  The Bayesian framework in \cite{asensio_ramos2007} seems best-suited to define confidence intervals of the returned parameters but is for now too computationally intensive to implement here}, spatial trends can be indicative of real changes. Field inclinations and azimuths are here given in the same local solar geometry whose reference axis is the solar vertical and polar axis (azimuth reference direction) points towards disk center.  For the fibrils exhibiting a $180^{\circ}$ ambiguity, we only display the solution that best matches the field direction at a height of 1.75 Mm within a current-free extrapolation of the photospheric magnetic field.  This extrapolation is computed from a $300''\times300''$ subregion of the full-disk SDO/HMI magnetogram centered on the active region and is based on the equations of \cite{alissandrakis1981} and \cite{gary1989}.  At this point, we use this only as an approximation as others means to disambiguate this $180^{\circ}$ Hanle ambiguity should be explored.  Of course, observations with better signal-to-noise will eliminate this ambiguity in some cases.

The fibril magnetic field strengths show a gradient towards lower field strengths outward from the sunspot down to 50 to 100 Gauss from 600 to 800 Gauss above the penumbra (magnetic field color table of Figure~\ref{fig:fibril_field_maps} saturates at 600 Gauss).  Inclinations of all fibril endpoints terminating within the sunspot show an increase in inclination consistent with a magnetic field turning downwards into the negative polarity sunspot.  These sunspot-rooted fibrils all display mostly horizontal field outside of the outer penumbral boundary.  The outer endpoint of these fibrils do not show a common pattern.  Fibrils \#32, \#27, and \#16 exhibit a turnover towards lower inclinations ($\theta_{B}<90^{\circ}$) at their outer endpoints giving the impression that these fibrils are anchored field loops rooted at one end in the sunspot and at the other in the nearby photosphere (see next section).  Fibrils numbered 2,6,7,9,and 11, however, remain mostly horizontal at their outer endpoints, with little indications whether the fields at this point turn upwards or downwards.  Fibril \#1 shows increases in inclinations with distance from the spot, but the inner endpoint is primarily horizontal and located outside the penumbral boundary.



\begin{figure*}
\epsscale{1.}
\plotone{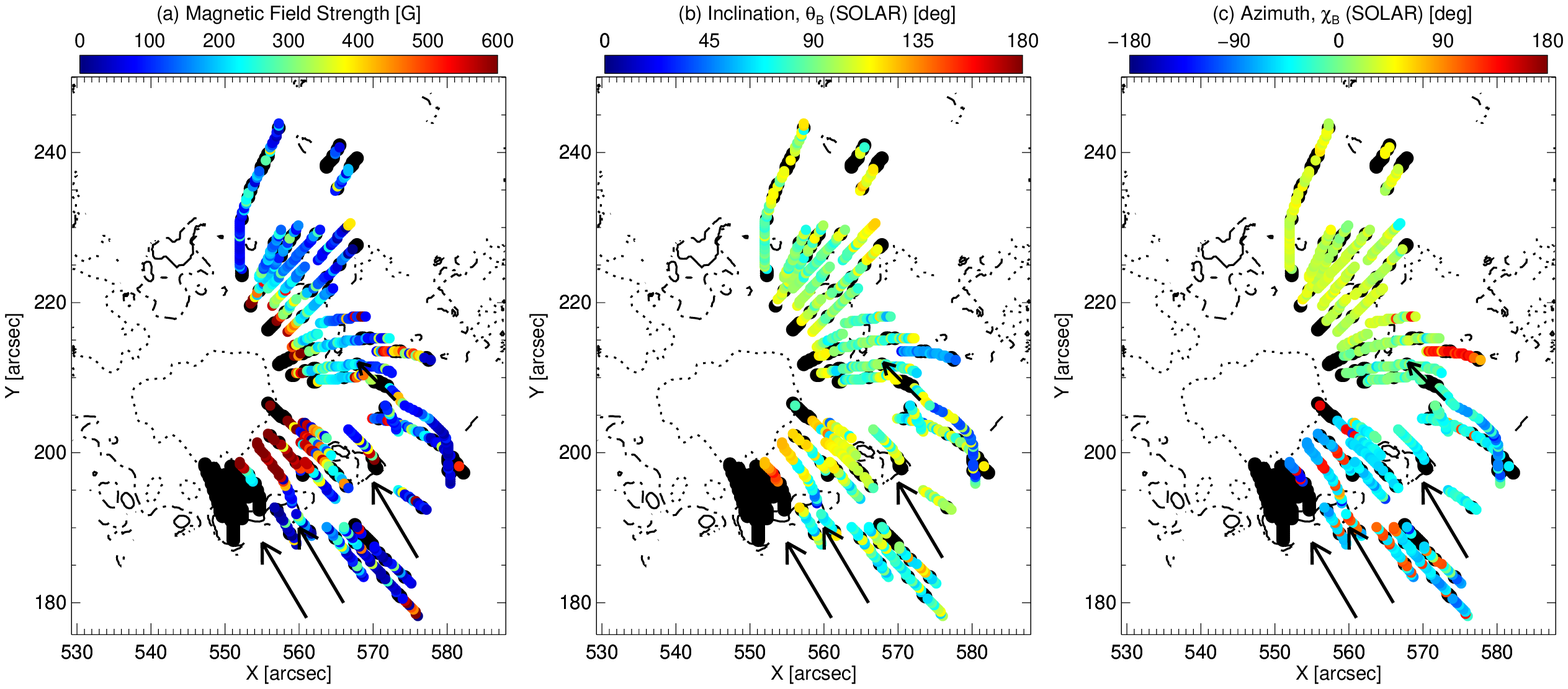}
\caption{Spatial maps of the magnetic field vector of superpenumbral fibrils inferred from the \ion{He}{1} triplet.  Black data points indicate locations for which the goodness-of-fit to the observed spectra is poor.  Contours of the photospheric magnetic field inclination are also given for values of $135^{\circ}$ (dotted), $90^{\circ}$ (dot-dashed), and $65^{\circ}$ (solid).  The inclination and azimuth values are given in the local solar reference frame.  Vector directions of many fibrils suffer from an $180^{\circ}$ Hanle ambiguity, which is here resolved with a potential field extrapolation of the photospheric magnetic field.  The black arrows indicate regions of plage with a polarity opposite w.r.t. the spot.  See the text for more details. \label{fig:fibril_field_maps}}
\end{figure*}

\begin{figure*}
\epsscale{1.}
\plotone{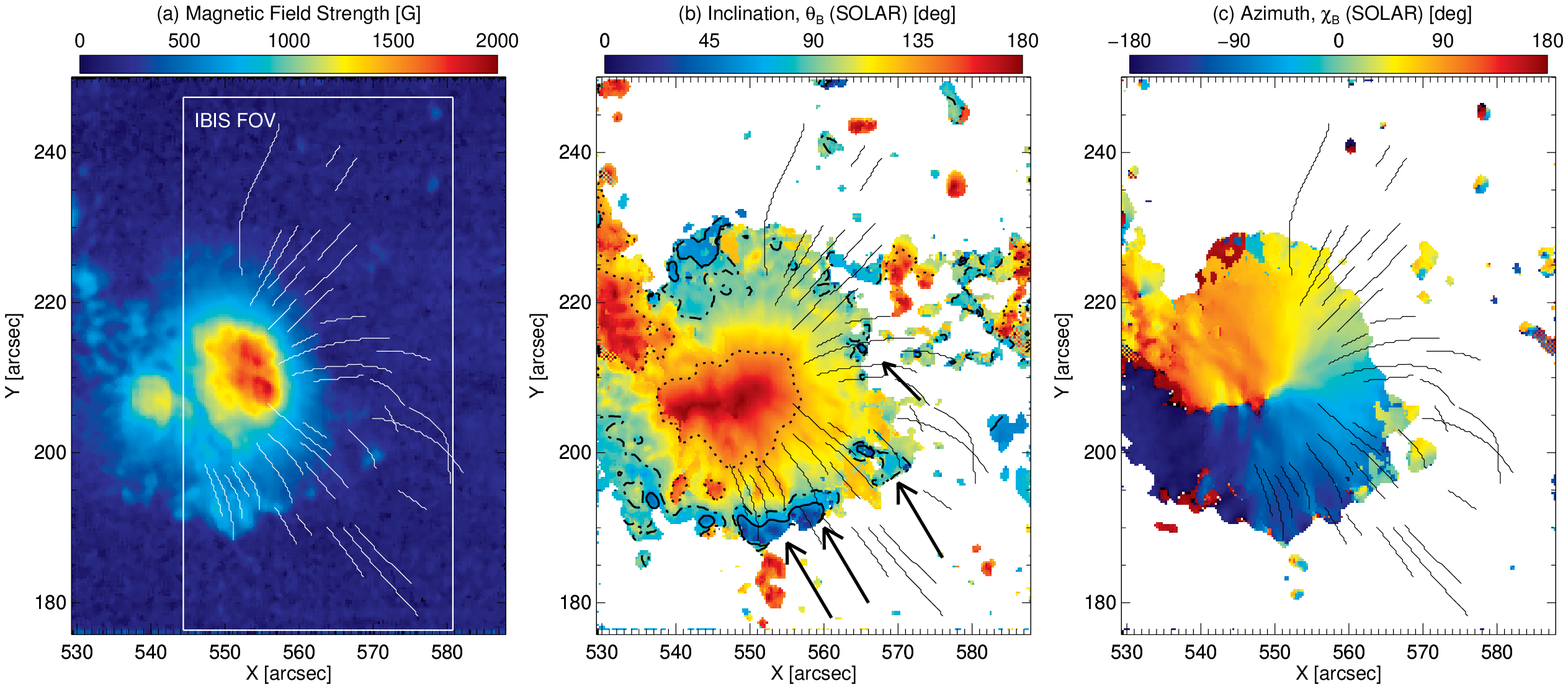}
\caption{Photospheric magnetic field vector in local solar coordinates for the entire FIRS field of view derived from a Milne-Eddington analysis of the Si I 10827.089 \mbox{\AA} spectral line.  Contours of the inclination angle are overplotted for the inclination plot at the same values as in Figure~\ref{fig:fibril_field_maps}.  The selected fibrils are also overplotted.  Arrows denote locations of the photospheric magnetic field with opposite polarity w.r.t. the sunspot umbra. \label{fig:photo_field}}
\end{figure*}

\subsection{The Underlying Photospheric Magnetic Field}\label{sec:photo_field} 

By comparing the fibril magnetic field vector and the photospheric vector magnetic field, we gain a more complete picture of the how chromospheric fibrils might be anchored at lower heights.  The Si I absorption line measured by FIRS at 10827.089 \mbox{\AA} ($g_{eff} = 1.5$) provides a good diagnostic of the photospheric magnetic field due to the Zeeman effect.  Furthermore, the Si I spectra are acquired strictly simultaneously with the \ion{He}{1} triplet at 10830 \mbox{\AA} for each spatial pixel in the FIRS scan.  As in \cite{bethge2012}, we use the Milne-Eddington (ME) inversion scheme implemented in the \textsc{HeLIx$^\textbf{+}$} inversion code \citep{lagg2004,lagg2007} to derive the vector magnetic field from the Si I line averaged over its formation height, which according to \cite{bard2008} is between 300 and 550 km above the solar surface for umbral-type and quiet solar atmospheres, respectively.  The derived azimuths include a $180^{\circ}$ ambiguity inherent in the transverse Zeeman-effect.  To resolve this ambiguity prior to transforming the magnetic field vector into local solar coordinates, we make use of the automated ambiguity-resolution code developed by \cite{leka2009}, which is based on the Minimum Energy Algorithm by \cite{metcalf1994}. 

The disambiguated, transformed vector magnetic field resultant from inversions of the Si I 10827.089 $\mbox{\AA}$ line is presented in Figure~\ref{fig:photo_field} in the same geometry as Figure~\ref{fig:fibril_field_maps}.  Contours are given for the solar inclination of this photospheric field at values of $135^{\circ}$ (dotted) ,  $90^{\circ}$ (dot-dashed), and $65^{\circ}$ (solid) and are plotted also in the maps of Figure~\ref{fig:fibril_field_maps}.  Azimuths in the chromospheric fibrils show general consistency with the azimuths of the penumbral filaments below.  Additionally, we plot the variation of the photospheric magnetic field and field-projected velocity (see Equation~\ref{eq:proj_vel}) directly below our representative fibrils in Figure~\ref{fig:photo_fibril}, except for \#31 which is above a region of the photospheric field not well represented by the combination of disambiguation and coordinate transformation. 

As described in section~\ref{sec:multi_wave}, the sunspot is trailed by opposite polarity plage.  Ahead (i.e. solar west) of the spot is a large area of plage matching the polarity of the sunspot.  Just south of the sunspot is a close-proximity area of flux with polarity opposite of the sunspot.  It is in this region that the outer endpoint of \#32 terminates providing evidence that the fibril is a closed field loop rooted in the sunspot on one end and in the opposite polarity flux on the other.  Fibril \#16 also shows this behavior.  Arrows in Figures~\ref{fig:fibril_field_maps} and~\ref{fig:photo_field} bring attention to areas of significant opposite signed flux.  Fibril \#15 exhibits behavior consistent with its one endpoint (close to the spot) rooted in this opposite polarity flux.  Fibrils numbered 2,6,7,9, and 11, discussed above, do not clearly terminate above opposite flux (see fibril 6 in Figure~\ref{fig:photo_fibril}).  The most prominent flux concentrations in this region are the leading plage with the same polarity as the sunspot.  Are these fibrils directed over the plage?  Or are they connected to unresolved footpoint fields below?  This becomes the same question as for internetwork flux discussed in the introduction.

\begin{figure*}
\epsscale{1.}
\plotone{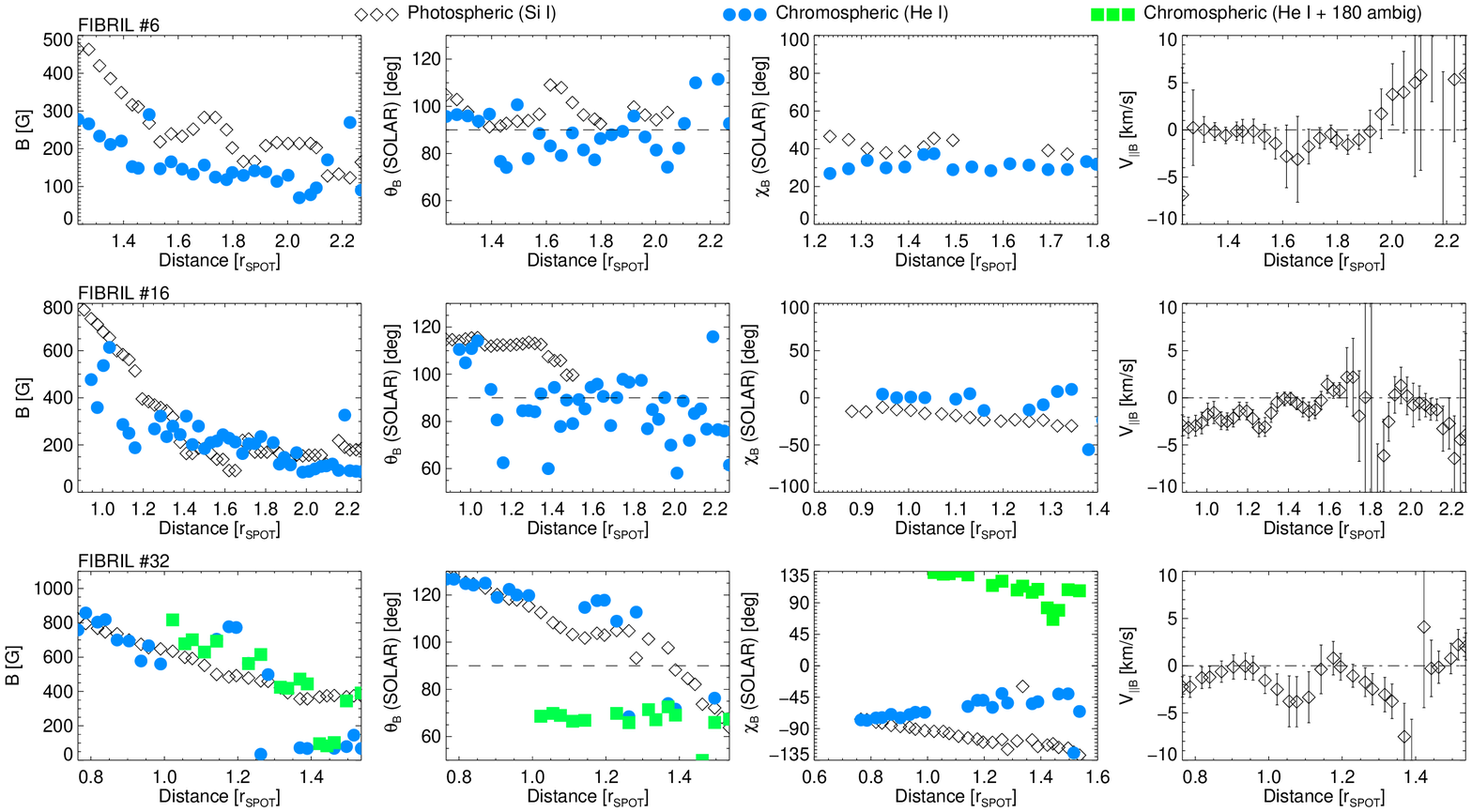}
\caption{Photospheric vector magnetic field resultant from a Milne-Eddington analysis of the Si I 10827.089 \mbox{\AA} spectral line shown for points directly below selected fibrils along the line-of-sight.  The magnetic field vector within the fibrils are overplotted.  The far right panel displays the component of the photospheric flow velocity along the photospheric magnetic field vector. (Color version available online) \label{fig:photo_fibril}}
\end{figure*}

\subsection{The Influence of Height of Inversion}\label{sec:infer_heights}

\begin{figure}
\epsscale{1.}
\plotone{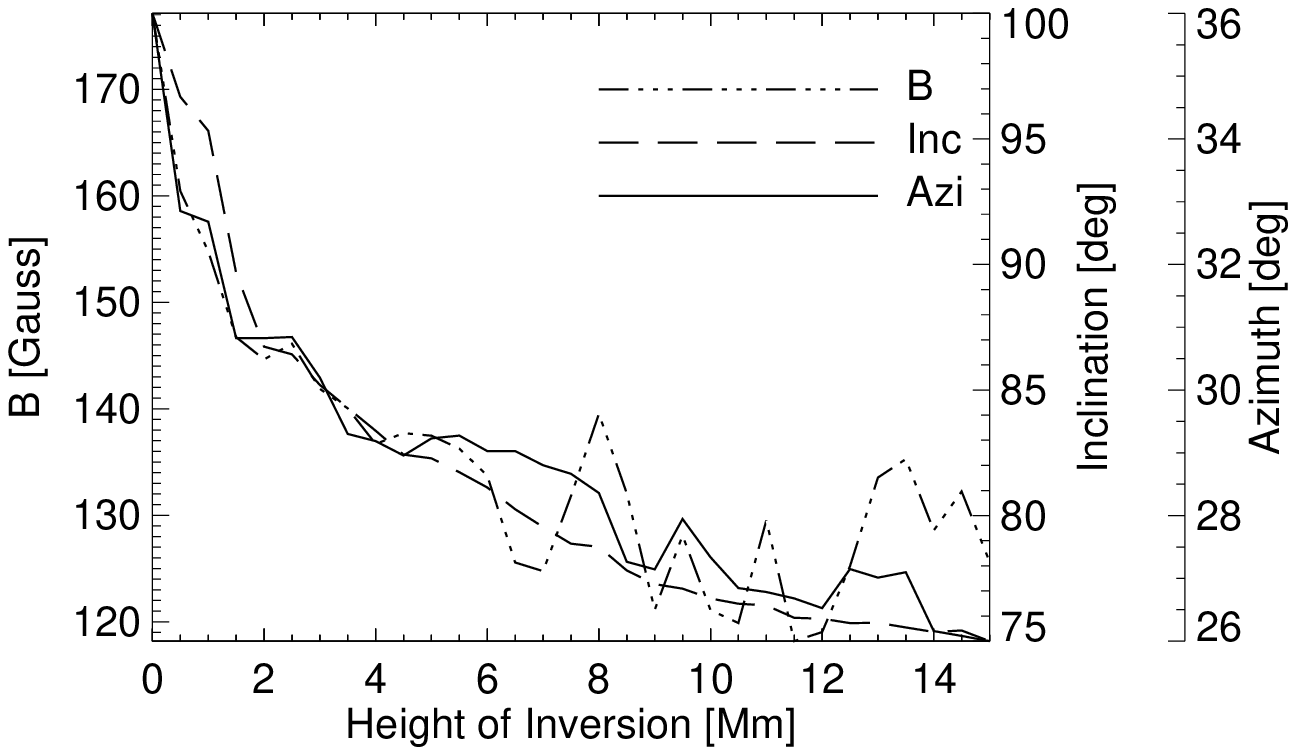}
\caption{The influence of the assumed height on the inversions of the \ion{He}{1} spectra within the superpenumbral fibrils.  The median magnetic field vector components of fibril no. 6 is shown as a function of the height of inversion.\label{fig:inv_height}}
\end{figure}

Our inversions do not take into account differences in the height of formation along individual fibrils or between multiple fibrils.  Rather, all inversions are carried out with the same assumed height of 1.75 Mm.  Ideally, one could use the spectra themselves to estimate the fibril heights either by letting height be a free parameter in addition to the magnetic field parameters \citep{merenda2011}, or by constraining the field direction ($\theta_{B}$ or $\chi_{B}$) by some other means (\refasen), such as by using the fibril direction itself.  Due to the influence of noise in our observations, we are unable to constrain the height in either of these ways.  As can be seen in Figures~\ref{fig:fib_para} and~\ref{fig:photo_fibril}, significant scatter exists in the determined field parameters along each fibril.  This scatter is used to estimate the error in the average fibril azimuths in Table~\ref{tbl:vis_vs_infer}.  Prohibitive as this scatter can be, we are confident that within the error, the fibrils are aligned with the magnetic field using the assumed height of 1.75 Mm.  In Figure~\ref{fig:inv_height}, we show the median field parameters of fibril \#6, tracking as it were the solution we found at a height of 1.75 Mm for all heights between 0 and 15 Mm.  This range extends well beyond what can be considered reasonable heights for fibril formation yet the change in the average inclination is only 20 degrees.  Figure 18 of \refasen{ }shows a similar dependence of the returned inclination with height.  Between 0 and 5 Mm, the variation in the inverted azimuthal angle is only 10 degrees.  Unfortunately, we cannot use this height influence to estimate the fibril formation height since the range in Figure~\ref{fig:inv_height} is on order of the error in the field parameters inferred at 1.75 Mm.  Thus, while we cannot further constrain the height, the influence of the height does not affect our primary results.


\section{Discussion}\label{sec:discussion}

\subsection{Relation of Fibrils to the Magnetic Field}

This paper has described the first vector magnetic field determinations within resolved superpenumbral fibrils.  Unlike observations in the \ion{Ca}{2} 8542 \mbox{\AA} spectral line \citep{delaCruz2011}, we find strong \ion{He}{1} linear polarization signatures originating from the superpenumbral canopy.  A heuristic intrepretion of these polarized spectra alongside advanced inversions from HAZEL lend support for field-aligned fibrils that are primarily horizontal ($\pm 20^{\circ}$) with respect to the solar surface.  We find little evidence for any misalignment of the thermal and magnetic structure of these fibrils and thus support extrapolation methods such as \cite{wiegelmann2008} and \cite{yamamoto2012}, which use the fibril direction to constrain the horizontal magnetic field direction in the chromosphere.  Yet, further comparisons need to be made between these extrapolations and actual measurements of the magnetic field strength and inclination, in particular due to the limitations of the force-free assumption in these extrapolations. 

A recent observation of a kink wave in an active region dynamic fibril \citep{pietarila2011} under assumptions regarding the unknown plasma density provided estimates of the fibril field strengths between and 100 and 350 Gauss.  We find magnetic field strengths less than 300 Gauss throughout the superpenumbra exterior to the penumbral boundary. The fibrils that extend into the sunspot penumbra exhibit a rise in field strengths.  Future time resolved measurement in \ion{He}{1} may grant a diagnostic of the unknown density when pairing field estimates with wave observations.

\subsection{Endpoint Connectivity of Fibrils}

Spatial trends in the inferred magnetic field inclination in both the chromosphere and photosphere are used here to study the 3D nature of superpenumbral fibrils.  We find evidence that most of the superpenumbral fibrils are rooted in the sunspot, as expected, and magnetic fields that are more vertically direction at that endpoint.  These fibrils become more horizontal with increased distance from the sunspot, especially as the fibrils cross the outer penumbral boundary.  A few of these fibrils turn over once again to connect in regions of oppositely directed flux in the photosphere.  Other do not show this behavior and remain nearly horizontal at their outer endpoints, which are located near plage of the same polarity as the sunspot.  In Foukal's 1971 picture, these fibrils might turn upwards into the upper atmosphere and connect with flux elsewhere.  We see no evidence for this in the fibril inclinations.  Furthermore, unlike H$\alpha$ and \ion{Ca}{2}, \ion{He}{1} can still sense cooler material at greater heights, suggesting the fibrils may be further extended in \ion{He}{1} than in H$\alpha$ if they indeed turn upwards.  The observations in Figure~\ref{fig:fov_maps} give little indication that this is happening; though these \ion{He}{1} maps suffer from poor temporal resolution due to the slit-scanning time.

It is not understood why we do not see a change in inclination at the outer endpoint of some fibrils.  We suggest projection effects may play a role and/or limited opacity of \ion{He}{1} at lower heights for some fibrils.  The later argument would require a thermodynamic difference between those fibrils with and without outer endpoint inclination changes.  Thus, we cannot conclude at this time that all the fibrils material represents closed field loops rooted just below both their endpoints, as argued by \cite{reardon2011} via visible constraints on the loop trajectory.  A way forward may be the study of the 3D velocity field in H$\alpha$ of \ion{Ca}{2} from IBIS, as in \cite{judge2010}.  \cite{ji2012} established for an arcade of low coronal loops a connectivity of fine-scaled neutral helium channels within intergranular lanes, as observed in very high resolution narrowband images of the \ion{He}{1} triplet at 10830 \mbox{\AA}.  These methods, in addition to further spectropolarimetric measurements, may aid in addressing the connectivity of fine-scaled internetwork and superpenumbral fibrils.

Nevertheless, the fibrils seen here to turn over at both ends correspond with an outer endpoint of stronger fields relative to the surrounding areas.  Fibril \#32, for example, is rooted at one endpoint in the strong sunspot and in $\sim375$ Gauss plage at the outer endpoint.  If the fibrils that exhibit no turn over (\#2,6,7,9,11) are in reality connected below, it must be within weaker flux elements ($B<200G$) or fine-scaled flux elements below the resolution of the observations. 

\subsection{Relation of Superpenumbral Flows to Magnetic Architecture}

Each observed fibril displays significant motion along its axis as measured with the \ion{He}{1} Doppler shift. The direction of the observed flow is consistent with inward-directed inverse Evershed flow.  Ultimately, we wish to study the thermo-magnetic properties of individual fibrils to understand the driving mechanisms of these flows and other observed phenomena.  Unfortunately, the poor temporal resolution of these FIRS observations limits our ability to comment on the temporal evolution of these flows.  However, we can comment on the magnetic architecture which hosts the flows.  

In siphon flow models of the chromospheric Evershed flow \citep{meyer1968,maltby1975,cargill1980}, the driving force is a gas pressure difference along the fibril caused by off-balanced magnetic pressure at the fibril footpoints.  \cite{bethge2012} examined a cool feature, interpreted as a loop, that displayed opposite signed LOS flows at its endpoints.  The magnetic flux difference at these endpoints could explain the observed flow magnitude in a siphon flow scenario.  \cite{bethge2012} argued that the material undergoes a deceleration at its stronger magnetic footpoint according to multi-wavelength observations.  Decelerated flow is a key component of siphon flow, in both the subsonic or shocked flow cases (see \cite{cargill1980}).  

In the outer footpoint, which hosts the upflow, \cite{bethge2012} showed LOS velocities of similar magnitude for \ion{He}{1} and \ion{Ca}{2} H 3968.5 \mbox{\AA}.  Unfortunately, due to the geometry involved, the acceleration of the up-flow is difficult to constrain without knowledge of the field geometry.  A common feature of the fibrils observed here is a lateral gradient of the LOS velocity along the fibrils.  Under the assumption that the flow is directed along the field, we derived the total magnitude of the flow (see Equation~\ref{eq:proj_vel} and Figure~\ref{fig:fib_para}).  Fibrils \#6 and \#16 showed weakly accelerating flows at their outer footpoints.  We estimate an acceleration of $\sim40m/s^{2}$ for the outer end of fibril \#16.  However, without finer determinations of the fibril inclination, we cannot at this time distinguish whether this acceleration is siphon driven or perhaps gravitationally driven.


\section{Concluding Remarks}\label{sec:conclusions}

We have employed high-spatial and high-spectral resolution spectropolarimetry of the \ion{He}{1} triplet at 10830 \mbox{\AA} to probe the local magnetic field vector in individual superpenumbral fibrils.  Key to this work has been the ability to achieve observations of high signal-to-noise and high polarization accuracy at high spatial resolution, which has come at the price of temporal resolution.  Yet, despite this limitation, the application of the advanced forward modeling and inversion techniques of \cite{asensio_ramos2008} yields several new inferences for the magnetic field vector within fibrils.  The primary conclusions of this work are as follows:
\begin{enumerate}
\item Superpenumbral fibrils do trace the magnetic field.  Despite the role of ambiguities in the determination of the magnetic field vector, \ion{He}{1} inversions nearly always return a solution whose projected field direction is consistent (generally within $\pm 10^{\circ}$) with the projected direction of the visible fibrils \par
\item The inner endpoint of superpenumbral fibrils hosts a detectable change in inclination as it turns into the sunspot where the fibrils are rooted. \par
\item Opposite-signed flux roots the outer endpoint of fibrils in at least two cases, but the connectivity of most of the fibrils is hard to establish.  If they are rooted below, as we suspect they are, they are connected to fine-scaled magnetic flux elements below the resolution of FIRS ($0.3''$).
\end{enumerate}

Perhaps none of these conclusions are that surprising.  We are confirming basic assumptions that have been made in the literature now for decades. Demonstrated here though is a powerful new means to measure both the macro-scaled and fine-scaled features of the chromospheric field with currently available instrumentation.  Yet, the relevant temporal scales of important upper atmospheric dynamics are still out of reach for chromospheric spectropolarimetry using the existing small-aperture solar facilities.  We stress the need for large-aperture facilities coupled to high-sensitivity (imaging-) spectropolarimeters both on the ground such as the Advanced Technology Solar Telescope \citep{rimmele2008}, and in new space missions, such as the Solar-C mission \citep{shimizu2011}.


\acknowledgments

We express our appreciation to the DST observers for their always devoted assistance with acquiring high-quality data, especially Doug Gilliam, Joe Elrod, Mike Bradford, and Crystal Hart.  Thanks to A. Tritschler and S. Jaeggli for their valuable expertise regarding IBIS and FIRS.  We also thank A. Asensio Ramos for help with HAZEL and R. Casini and P. Judge for valuable discussions.  FIRS has been developed by the Insitute for Astronomy at the University of Hawai'i jointly with the NSO, and was funded by the National Science Foundation Major Research Instrument program, grant number ATM-0421582.  .  IBIS was constructed by INAF/OAA with contributions from the Universities of Florence and Rome, and is operated with support of the NSO.  We acknowledge the courtesy of NASA/SDO and the AIA and HMI science teams for providing high quality data.

{\it Facilities:} \facility{Dunn Solar Telescope}


\end{document}